\documentclass[12pt]{article}

\usepackage{graphics,psfrag,rotating,amssymb}
\usepackage{cite} 
\usepackage{epsfig}
\usepackage{amsmath}
\newenvironment{Eqnarray}%
     {\arraycolsep 0.14em\begin{eqnarray}}{\end{eqnarray}}

\makeatletter
\@addtoreset{equation}{section}
\def\theequation{\thesection.\arabic{equation}}
\makeatother

\newcommand{\nn}{\nonumber}

\newcommand{\be}{\begin{equation}}
\newcommand{\ee}{\end{equation}}
\newcommand{\bea}{\begin{Eqnarray}}
\newcommand{\eea}{\end{Eqnarray}}

\def\lsim{\mathrel{\raise.3ex\hbox{$<$\kern-.75em\lower1ex\hbox{$\sim$}}}}
\def\gsim{\mathrel{\raise.3ex\hbox{$>$\kern-.75em\lower1ex\hbox{$\sim$}}}}
\def\ifmath#1{\relax\ifmmode #1\else $#1$\fi}
\def\beq{\begin{equation}}
\def\eeq{\end{equation}}
\def\beaa{\begin{array}}
\def\eeaa{\end{array}}

\def\ls#1{\ifmath{_{\lower1.5pt\hbox{$\scriptstyle #1$}}}}
  
\renewcommand{\thefootnote}{\fnsymbol{footnote}}

\begin{document}
\hfill{} 
\begin{tabular}{l} 
MPP-2003-134\\ 
FTUAM-03-23 \\
IFT-UAM/CSIC-03-52 \\
\vspace*{0.2cm}
\end{tabular} 
\begin{center}
{\Large\bf
SUSY-electroweak one-loop contributions to \\[0.3cm] 
Flavour-Changing 
Higgs-Boson Decays} \\[1.3cm] 
{\large 
A.~M.~Curiel$^{\,\scriptstyle{a}}$,
M.~J.~Herrero$^{\,\scriptstyle{a}}$, W.~Hollik$^{\,\scriptstyle{b}}$, 
F.~Merz$^{\,\scriptstyle{b}}$  and S.~Pe{\~n}aranda$^{\,\scriptstyle{b}}$
~\footnote{electronic addresses:
ana.curiel@uam.es, maria.herrero@uam.es,
hollik@mppmu.mpg.de, merz@mppmu.mpg.de, siannah@mppmu.mpg.de}
}\\[7pt]
{\it a}: {\it Departamento de F\'{\i}sica Te\'{o}rica,
   Universidad Aut\'{o}noma de Madrid,
   Cantoblanco, 28049 Madrid, Spain.}\\[1pt]
{\it b}: {\it Max-Planck-Institut f\"ur Physik, F\"ohringer Ring 6, 
D--80805 Munich, Germany}
\\[1cm]

\begin{center}
{\bf Abstract}
\end{center}
\end{center}

The SUSY-EW one-loop quantum contributions to flavour-changing MSSM 
Higgs-boson decays into $b \bar s$ and $s \bar b$ are computed and
discussed. We use the full diagrammatic approach that is valid for all 
$\tan \beta$ values and do not rely on the mass-insertion approximation 
for the characteristic flavour-changing parameter. We analyze in full 
detail the dependence of these flavour-changing partial widths on  
all the relevant MSSM parameters and also
 study the non-decoupling behaviour of these widths with the SUSY mass 
parameters. We find that these contributions are sizable as compared to the SM 
ones, and together with the SUSY-QCD contributions
they can be very efficient as an
indirect method in the future search for Supersymmetry.
  
\vfill
\clearpage

\renewcommand{\thefootnote}{\arabic{footnote}}
\setcounter{footnote}{0}

\pagestyle{plain}
\section{Introduction}
\label{chap.introduction}
 
Rare processes involving Flavour Changing Neutral Currents (FCNC) provide an 
extremely useful tool to investigate new physics beyond the Standard 
Model (SM). The strong suppression of FCNC in the SM is due to the absence of
tree-level contributions and the smallness of the loop-contributions. 
The later, being a consequence of the GIM-cancellation
mechanism~\cite{GIM}, naturally enhances the sensitivity of Flavour 
Changing (FC) processes to possible non-standard phenomena. 
Among the various possibilities, the
 Higgs-mediated FCNC processes involving down-type quarks are particularly
interesting since, within the SM, they are additionally suppressed by the 
smallness of the down-type Yukawa couplings.

The Minimal Supersymmetric Standard Model (MSSM)~\cite{HaberKane} 
provides a natural framework 
where such scalar FC interactions could be significant if the  
soft SUSY-breaking mass terms contain some 
non-diagonal structure in flavour space. In the
minimal-flavour-violation scenario of the MSSM, squarks are assumed to be
aligned with the corresponding quarks.
Flavour violation in this case  originates from the
Cabibbo-Kobayashi-Maskawa (CKM) matrix as the only source,
and proceeds  via loop-contributions, in analogy to the SM. 
Therefore, its size
is expected to be very small. However, in the more general scenarios that
include misalignment between the quark and squark sectors, sizeable
contributions to FCNC processes are expected to occur.
It is indeed the case
of the radiative corrections from SUSY-loops in the context of $B$-meson 
physics that come with factors of $\tan \beta$, the ratio of the vacuum 
expectation values of the two MSSM Higgs doublets. These $\tan \beta$-enhanced
SUSY radiative corrections have been studied in a number of processes, 
including $B^0-\bar B^0$ mixing~\cite{toharia,newchanko}, 
leptonic $B$ meson decays~\cite{kolda,oldchanko,alemanes,newisidori}, and
$b \rightarrow s \gamma$ 
decays~\cite{bstanbeta,bsgamma,leadingbsg,gambinobsg}, 
and have been found to be large. In order not to be in conflict with 
present experimental data, these in turn imply some restrictions on 
the parameters measuring the size of flavor mixing in the squark sector 
which, as we have said, is mainly produced from quark--squark misalignment~\cite{newisidori,9604387,ciuchini,Endo}.    

Other FCNC processes of interest are related to Higgs physics and are 
also very sensitive to supersymmetric quantum 
effects~\cite{Maria,dedes,Demir,radFCNC}. In particular, the neutral 
MSSM Higgs-boson decays into $b  \bar s$ and $s \bar b$ have been proven to be 
generated quite efficiently from squark--gluino loops if quark--squark
misalignment is assumed~\cite{Maria}. These SUSY-QCD loop contributions have
the peculiarity of being non-vanishing even in the limit of very heavy
SUSY particles and, in addition, they are enhanced by large
$\tan \beta$ factors. Such a  non-decoupling behaviour of SUSY particles 
in the Flavor Changing Higgs Decays (FCHD) can be of special
interest for indirect SUSY searches at future colliders, as the forthcoming 
LHC and a next $e^+e^-$ linear collider, in particular if the
SUSY particles turn out to be too heavy to be produced directly. The large rates 
found for the SUSY-QCD contributions to the Higgs partial decay widths into
$b \bar s$ and $s \bar b$~\cite{Maria}, as well as to the effective 
FC Higgs couplings to quarks~\cite{kolda,oldchanko,alemanes,newisidori,dedes,Demir}, 
are indeed quite encouraging.

In this paper we  complete the  genuine SUSY quantum effects in FCHD 
by the computation of the SUSY electroweak (SUSY-EW) one-loop contributions 
to Higgs-boson decays into  $b \bar s$ and $s \bar b$. 
The dominant SUSY-EW radiative corrections to the effective Higgs-boson 
couplings to quarks have been computed recently in the mass-insertion 
approximation, including both misalignment and CKM induce
effects~\cite{Demir}. Here we will perform an exact computation 
of the complete SUSY-EW one-loop contributions from  
squark--chargino and squark--neutralino loops to the flavour-nondiagonal 
decay rates of the three neutral MSSM Higgs bosons
and  compare them with the SUSY-QCD contributions.
Since no approximation is used, our computation is valid for all values of the
characteristic parameter measuring the squark-mixing strength and for all 
$\tan \beta$ values.  Furthermore, we will explore in detail the dependence of
these SUSY-EW quantum contributions on the MSSM parameters, 
and we will study their
behaviour in the large sparticle-mass limit. Both the numerical results
and the asymptotic analytical formulas, showing the non-decoupling 
behaviour, can be of particular interest for future Higgs-physics studies at
next-generation colliders.

The paper is organized as follows. Section 2 describes the basis for FC 
interactions in the SUSY-EW sector of the MSSM. The
computation of the SUSY-EW one-loop contributions to the Higgs-$b$-$s$ 
form factors and Higgs decay widths into $b \bar s$ and $s \bar b$ are outlined
in section 3. The numerical analysis of the FCHD rates and a detailed
discussion of the dependence on the relevant MSSM parameters are included in
section 4, and the
large sparticle-mass limit is studied in
section 5. A set of useful and compact formulas valid in this limit is also 
derived; it is listed in the Appendix, 
together with other details of the computation.

\section{Flavour-changing interactions in the MSSM}
\label{FC_MSSM} 

In the MSSM there are two sources of FC phenomena. The first one 
is common
to the Standard Model case and is due to mixing in the quark sector. It is 
produced by 
the different rotation in the $d$- and $u$-quark sectors, and its strength 
is driven by the off-diagonal CKM-matrix elements. This mixing produces 
FC electroweak
interaction terms involving  charged currents and, in particular,  
supersymmetric electroweak  
interaction terms of the chargino--quark--squark type, 
which are of interest to this work. 
The second source of FC phenomena is due to the possible 
misalignment between the rotations that diagonalize the quark and 
squark sectors.  When the squark-mass matrix is
expressed in the basis where the squark fields are parallel to the
quarks (the super-CKM basis), it is in general
non-diagonal in flavour space. This quark--squark misalignment produces 
new FC terms in
neutral-current as well as in charged-current interactions. 
In the case of the SUSY-QCD sector, the FC interaction terms 
involve neutral currents of the gluino--quark--squark type, and their effects 
on FCHD have been studied in ref.~\cite{Maria}. Here we focus on
the SUSY-EW interaction terms generating FC phenomena, in particular on those  
of the neutralino--quark--squark and chargino--quark--squark type.
The first one appears exclusively due to
quark--squark misalignment, as in the SUSY-QCD case, whereas the second 
one receives contributions from both sources, quark--squark misalignment and
CKM mixing. 

We assume here that the non-CKM
squark mixing is significant only for transitions between the 
third- and second-generation
squarks, and that there is only LL mixing, given by a similar
ansatz as in~\cite{AnsatzSher} where it is proportional to the product of the
 SUSY masses involved. This assumption is theoretically well motivated  by the
  radiatively induced flavour off-diagonal 
squark squared-mass entries via RGE from high energies down to the electroweak 
scale~\cite{Hikasa}. These RGE predict that the flavour changing LL entries scale with the square of the soft-SUSY
 breaking masses, in contrast with the LR (or RL) and the RR entries that scale
 with one or zero powers, respectively. Thus,
 the hierarchy $LL>>LR \,, RL >> RR$ is usually assumed. These same estimates 
also indicate that the LL entry for the mixing between the second and third generation squarks is 
the dominant one due to the larger quark mass factors involved. On the 
other hand, the LR and RL entries are experimentally more constrained, mainly 
by $b \to s \gamma$ data~\cite{ciuchini}. With the previous assumption, the squark squared-mass
 matrices in the ($\tilde c_L$,$\tilde c_R$,$\tilde t_L$,$\tilde t_R$) and 
($\tilde s_L$,$\tilde s_R$,$\tilde b_L$,$\tilde b_R$)
 basis, respectively, can be written as follows,
\begin{eqnarray}
M^2_{\tilde u} =\left\lgroup 
         \beaa{llll}
          M_{L,c}^2  &   m_c X_c & \lambda_{LL} M_{L,c} M_{L,t}& 0\\
           m_c X_c   &  M_{R,c}^2  & 0  &  0\\
          \lambda_{LL} M_{L,c} M_{L,t} & 0 & M_{L,t}^2  &   m_t X_t\\
          0 & 0 &  m_t X_t &  M_{R,t}^2 
\eeaa
         \right\rgroup ,
\label{eq.usquarkmass}
\end{eqnarray}
   \begin{eqnarray} 
M^2_{\tilde d} =\left\lgroup 
         \beaa{llll}
          M_{L,s}^2  &   m_s X_s & \lambda_{LL} M_{L,s} M_{L,b}& 0\\
           m_s X_s   &  M_{R,s}^2  & 0  &  0\\
          \lambda_{LL} M_{L,s} M_{L,b} & 0 & M_{L,b}^2  &   m_b X_b\\
          0 & 0 &  m_b X_b &  M_{R,b}^2
\eeaa
         \right\rgroup
\label{eq.dsquarkmass}
\end{eqnarray}
where
\begin{eqnarray}
M_{L,q}^2 &=& M_{\tilde Q,q}^2 +m_q^2 + \cos2\beta (T_3^{q}-Q_q s_W^2)m_Z^2\, , \nn \\
M_{R,(c,t)}^2 &=& M_{\tilde U,(c,t)}^2 +m_{c,t}^2 + \cos2\beta Q_t s_W^2 m_Z^2\, , \nn \\
M_{R,(s,b)}^2 &=& M_{\tilde D,(s,b)}^2 +m_{s,b}^2 + \cos2\beta Q_b s_W^2 m_Z^2\, , \nn \\
X_{c,t} &=& A_{c,t} - \mu \cot \beta \, , \nn \\
X_{s,b} &=& A_{s,b} - \mu \tan \beta \, ;
\label{eq.squarkparam}
\end{eqnarray}
$m_q$, $T_3^q$, $Q_q$ are the mass, isospin and electric charge 
of the quark $q$;  $m_Z$ is the $Z$ boson mass, and $s_W = \sin\theta_W$ contains
the electroweak mixing angle $\theta_W$. 
The relevant MSSM parameters in the SUSY-EW sector are, 
as usual, the soft SUSY-breaking EW gaugino masses, $M_1$ and $M_2$, 
the $\mu$-parameter, the soft
SUSY--breaking scalar masses $M_{\tilde Q}$, $M_{\tilde U}$, $M_{\tilde D}$, 
and the soft SUSY--breaking trilinear parameters, $A_q$. 
Owing to the $SU(2)_L$ invariance, $M_{\tilde Q,c} = M_{\tilde Q,s}$ 
and $M_{\tilde Q,t} = M_{\tilde Q,b}$. For simplicity, we have assumed 
the soft-breaking trilinear matrices to be diagonal in flavour space.

In our parametrization (\ref{eq.usquarkmass}), (\ref{eq.dsquarkmass})
of flavour mixing in the squark sector, 
there is only one free parameter, $\lambda_{LL}$, that characterizes the
flavour-mixing strength.  For the sake of simplicity, 
 we assume the same $\lambda_{LL}$ parameter 
 in the $\tilde t-\tilde c$  and  $\tilde b - \tilde s$ sectors,
 writing $\lambda \equiv \lambda_{LL}$ from now on for a simpler notation.
 Obviously, the choice
 $\lambda=0$ represents the case of zero  flavour mixing.

 In order to diagonalize the two
$4 \times 4$ squark-mass matrices given above, two $4 \times 4$ matrices, 
$R^{(u)}$ for the $up$-type squarks and 
$R^{(d)}$ for the $down$-type squarks, 
are needed.  Diagonalization yields the squark-mass eigenvalues
and eigenstates depending on $\lambda$. This dependence
has been studied in~\cite{Maria}; typically, two of the eigenvalues 
are weakly dependent on $\lambda$, with mass values very close to the case 
with $\lambda = 0$, and for the other two, one grows with $\lambda$ and the other decreases with it. For $\lambda=0$, one recovers the usual
pairs of physical flavour-diagonal squarks, ($\tilde b_1, \tilde b_2$),  
($\tilde s_1, \tilde s_2$), ($\tilde t_1, \tilde t_2$), 
and ($\tilde c_1,\tilde c_2$).

The following paragraph 
presents the SUSY-EW interaction terms that are responsible
for flavour-changing neutral Higgs-boson decays. 
We will write them in the mass-eigenstate basis. 

The chargino--quark--squark interactions are described 
by the  interaction Lagrangian
\begin{eqnarray}
\mathcal{L}_{\tilde \chi_j^- d \tilde u_{\alpha} } &=& 
-g\, \bar{d} \left[ A_{L \alpha j}^{(d)} P_L + 
A_{R \alpha j}^{(d)} P_R \right] \tilde \chi_j^- \tilde u_{\alpha} + h.c., 
\end{eqnarray} 
where $d$ can be either a $b$ or $s$ quark; the chargino index is $j=1,2$ for the
two physical states, and the
squark index is $\alpha=1,2,3,4$, representing the four physical squark
states. 
$g$ denotes the $SU(2)_L$ gauge coupling, and the coefficients
$A_{L \alpha j}^{(d)}$ and 
$A_{R \alpha j}^{(d)}$ are listed in Appendix~A. 
They include the two above commented sources of flavour-changing vertices, 
quark--squark misalignment and CKM mixing.

The neutralino--quark--squark FC interactions are described by
\begin{eqnarray}
\mathcal{L}_{\tilde \chi_a^0 b \tilde d_{\alpha} } &=& 
-g\bar{b} \left[ B_{L \alpha a}^{(b)} P_L + 
B_{R \alpha a}^{(b)} P_R \right] \tilde \chi_a^0 \tilde d_{\alpha} + h.c.\,,\nonumber\\ 
\mathcal{L}_{\tilde \chi_a^0 s \tilde d_{\alpha} } &=& 
-g\bar{\tilde \chi}_a^0 \left[ E_{L \alpha a}^{(s)} P_L + 
E_{R \alpha a}^{(s)} P_R \right]  s \tilde d_{\alpha}^* + h.c.\,, 
\end{eqnarray} 
where the neutralino index is $a=1,2,3,4$ for the four physical states
(the squark index is again $\alpha=1,2,3,4$), 
and the expressions for the coefficients  $B_{L \alpha a}^{(b)}$, 
$B_{R \alpha a}^{(b)}$, $E_{L \alpha a}^{(s)}$,
$E_{R \alpha a}^{(s)}$ are listed in Appendix A. 
Notice that FC effects originate 
only from quark--squark misalignment in this case.

The chargino--quark--squark and neutralino--quark--squark interactions
specified so far induce flavour-changing Higgs-boson decays like  
$H_x\rightarrow b \bar s$ and  $H_x\rightarrow s \bar b$ ($H_x=h^0,H^0,A^0$)
via electroweak one-loop contributions involving virtual 
squarks and charginos or neutralinos. 
We will study those effects in full detail in the 
forthcoming sections~\ref{chap.SUSY-EW_cont},~\ref{chap.phenomenology} 
and~\ref{chap.decoupling}.

For completeness, we list also the remaining interaction terms
which are not of the flavour-changing type,
but enter the Higgs-decay matrix elements and are thus  
relevant for the present work. 
These are the Higgs--quark--quark, Higgs--squark--squark, Higgs--chargino--chargino,
and Higgs--neutral\-ino--neutral\-ino interactions, reading
\begin{eqnarray}
\mathcal{L}_{H_xqq} &=& 
-g H_x\bar{q} \left[ S_{L,q}^{(x)} P_L + S_{R,q}^{(x)} P_R \right]  q \,\,,\nonumber\\ 
\mathcal{L}_{H_x \tilde q_{\alpha} \tilde q_{\beta}}&=&
-iH_x \left[ g_{H_x\tilde u_{\alpha} \tilde u_{\beta}} \tilde u_{\alpha}^* 
\tilde u_{\beta}
+ g_{H_x\tilde d_{\alpha} \tilde d_{\beta}} \tilde d_{\alpha}^* \tilde d_{\beta}
\right]\,\,,\nonumber\\
\mathcal{L}_{H_x \tilde \chi_i^- \tilde \chi_j^-}&=&
- g H_x\bar{\tilde{\chi}}_i^- 
\left[ W_{Lij}^{(x)}P_L+ W_{Rij}^{(x)}P_R \right] \tilde{\chi}_j^-\,\,,\nonumber\\
\mathcal{L}_{H_x \tilde \chi_a^0 \tilde \chi_b^0}&=&
- \frac{g}{2} H_x\bar{\tilde{ \chi}}_a^0 
\left[ D_{Lab}^{(x)}P_L+ D_{Rab}^{(x)}P_R \right] \tilde {\chi}_b^0\,\,,
\end{eqnarray}
($\alpha, \beta =1,2,3,4$, $i,j=1,2$, $a,b=1,2,3,4$), 
with the coefficients  
$S_{L,q}^{(x)}$, $S_{R,q}^{(x)}$, $W_{Lij}^{(x)}$, 
$W_{Rij}^{(x)}$, $D_{Lab}^{(x)}$, $D_{Rab}^{(x)}$ specified in Appendix A. 
The couplings $g_{H_x\tilde u_{\alpha} \tilde u_{\beta}}$ and 
$g_{H_x\tilde d_{\alpha} \tilde d_{\beta}}$ can be found in ref.~\cite{Maria}.

\section{Generating flavour-changing Higgs decays} 
\label{chap.SUSY-EW_cont}

We present in this section the computation of 
the loop-induced flavour-changing
neutral Higgs-boson decays 
into second and third generation quarks, 
$H_{x} \to b \bar s, s \bar b$, for $H_{x}=h^0,H^0,A^0$. 
We  focus on the one-loop contributions 
following from the SUSY electroweak sector,
i.e.\ on squark--chargino/neutralino loops. 
One-loop contributions from the 
SUSY-QCD sector have been analyzed in~\cite{Maria}. We follow in this
paper the notation introduced in~\cite{Maria}. 

\begin{figure}[t]
\begin{center}
\includegraphics[height=1.05in,clip=]{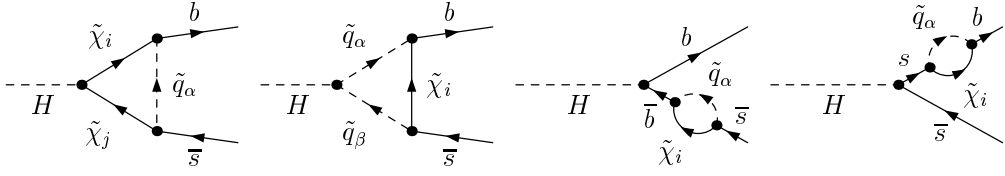}
\end{center}
\vspace*{-0.3cm}
\caption{Generic one-loop diagrams for 
  squark--chargino/neutralino 
   contributions to the decay $H \to b \bar s$. 
   $\tilde \chi \equiv \tilde \chi^{-}, \tilde
  \chi^{0}$
and $\tilde q \equiv \tilde u, \tilde d$, accordingly, with indices
$\alpha, \beta =1, ..., 4$ for squarks,  $i, j =1,
  ..., 4$ for neutralinos, and $i,j =1,2$ for charginos. 
These diagrams will be refered to as $a_C$, $b_C$, $c_C$ 
and $d_C$ ($a_N$, $b_N$, $c_N$, $d_N$) for charginos (neutralinos), 
respectively, from left to right.}
\label{fig.diagrams}
\end{figure}

For the partial decay widths, the one-loop matrix element
for each decay process $H \to q \bar q'$,
in compact form, can be written as follows, 
\begin{equation}
i F = -i g\,  \bar u_q  
 \left[ F_L^{qq'} (H) P_L + F_R^{qq'} (H) P_R \right]  v_{q'} H\, ,
\end{equation}
with the projectors $P_{L,R} =1/2 (1\mp \gamma_{5})$. 
$F_L$ and $F_R$ are the 
form factors for each chirality projection $L$ and $R$, respectively.
$F_L$ and $F_R$ follow from the explicit calculation of 
the vertex and non-diagonal self-energy diagrams,
depicted generically for the $b \bar s$ case in
Fig.~\ref{fig.diagrams}. Similar diagrams appear in the $s \bar b$ case. 

For $q=b$, $q'=s$, the form factors read explicitly, 
with the notation $F_{L,R}^{(x)} = F_{L,R}^{bs}(H_x)$ 
for a given Higgs boson $H_x$,
\begin{eqnarray}\hspace*{-1.2cm}
&&\hspace*{-0.5cm}F_L^{(x)} = - \frac{g^2}{16 \pi^2}\left[\left(B_0 + m_{\tilde d_{\alpha}}^2 C_0+m_s^2 C_{12}+m_b^2 (C_{11} - C_{12})\right)\,
\kappa_{L 1}^{x,\,\tilde \chi^0} \right.\nn \\
&&+m_b m_s \left(C_{11}+C_0\right)\kappa_{L 2}^{x, \tilde \chi^0}+m_b m_{\tilde \chi _b^0} \left(C_{11}-C_{12}+C_0\right)\kappa_{L 3}^{x, \tilde \chi^0}\,+ m_s m_{\tilde \chi _b^0} C_{12}\,\kappa_{L 4}^{x,\tilde \chi^0}\nn \\
&& +m_b m_{\tilde \chi _a^0} \left(C_{11}-C_{12}\right)\kappa_{L 5}^{x,\tilde \chi^0} + m_s m_{\tilde \chi _a^0} \left(C_{12}+C_0\right)\kappa_{L 6}^{x,\tilde \chi^0}+ m_{\tilde \chi _a^0} m_{\tilde \chi _b^0}C_0\,\kappa_{L 7}^{x,\tilde \chi^0} \left.\right] \nn \\
&&- \frac{igg_{H_x\tilde d_\alpha \tilde d_\beta}}{16\pi^2}\left[-m_b(\tilde C_{11}-\tilde C_{12})\,\iota_{L 1}^{x,\tilde \chi^0}-m_s \tilde C_{12}\, \iota_{L 2}^{x,\tilde \chi^0}+m_{\tilde \chi_a^0}\tilde C_0\,\iota_{L 3}^{x,\tilde \chi^0}\right]\nn \\
&&- \frac{S_{L,b}^{(x)}}{m_s^2-m_b^2}
\left[m_s^2 \Sigma_L^{\tilde \chi^0} (m_s^2)+
m_s m_b \Sigma_{Rs}^{\tilde \chi^0}(m_s^2)\right.
+ \left. m_b\left(m_s \Sigma_R^{\tilde \chi^0} (m_s^2)
+m_b \Sigma_{Ls}^{\tilde \chi^0} (m_s^2)\right)\right]\nn \\
&&- \frac{S_{L,s}^{(x)}}{m_b^2-m_s^2}\left[m_b^2 \Sigma_R^{\tilde
    \chi^0} (m_b^2)+m_b^2 \Sigma_{Rs}^{\tilde \chi^0}(m_b^2)\right.
+ m_s\left(m_b \Sigma_L^{\tilde \chi^0} (m_b^2)+m_b\Sigma_{Ls}^{\tilde \chi^0} (m_b^2)\right)\left.\right]\nn \\[0.3cm]
&&+ \;\; (\tilde \chi^0 \rightarrow \tilde \chi^-,\: a
\rightarrow i, \:b\rightarrow j )\,,
\label{formfactorLbs}
\end{eqnarray}
where summation over the various squark and chargino/neutralino indices is
to be understood.
For the 2-point and 3-point integrals, $B_0$,
$C_0$, $C_{11}$, $C_{12}$, taken from ref.~\cite{Hollik},
we have introduced an abbreviation for the arguments 
such that for neutralinos, $B=B(M_{H_x}^2,m_{\tilde \chi_a^0}^2, m_{\tilde \chi_b^0}^2)$,
$C=C(m_b^2, M_{H_x}^2, m_s^2, m_{\tilde d_{\alpha}}^2, m_{\tilde \chi_a^0}^2, m_{\tilde \chi_b^0}^2)$ and 
$\tilde C=C(m_b^2, M_{H_x}^2, m_s^2, m_{\tilde \chi_a^0}^2, m_{\tilde d_{\alpha}}^2, m_{\tilde d_{\beta}}^2)$, and similarly for charginos but replacing $\tilde \chi_{a,b}^0$ by $\tilde \chi_{i,j}^-$ and $\tilde d_{\alpha,\beta}$ by $\tilde u_{\alpha,\beta}$.
The expression for the right-handed form factor
$F_R^{(x)}$ can be obtained from $F_L^{(x)}$
by replacing $L\leftrightarrow R$ in all the terms
given in eq.~(\ref{formfactorLbs}). 
The definitions of the $\kappa$ and $\iota$ factors and of $S_q^{(x)}$
can be found in Appendix A. 

Besides the vertex integrals, the form factors
contain contributions from the flavour-non-diagonal
2-point functions, denoted by 
$\Sigma^{\tilde \chi}$ for $\tilde \chi = \tilde{\chi}^0, \tilde{\chi}^-$.  
Eq.~(\ref{formfactorLbs}) contains the 
scalar coefficients in the following 
Lorentz-decomposition of the non-diagonal self-energies, 
\begin{eqnarray} 
\Sigma^{\tilde \chi} (k) &=& k{\hspace{-6pt}\slash}\, \Sigma_L^{\tilde \chi} (k^2) P_L + 
 k{\hspace{-6pt}\slash} \,
\Sigma_R^{\tilde \chi} (k^2) P_R + m_b \left[ \Sigma_{Ls}^{\tilde \chi}(k^2) P_L + 
\Sigma_{Rs}^{\tilde \chi} (k^2) P_R \right]\, . 
\end{eqnarray}
The corresponding expressions can be found in Appendix~A.

The results of this computation have been obtained in two independent
ways, one without and one with the support of {\it FeynArts} and 
{\it FormCalc}~\cite{hahn}, and agreement was found. Thereby, the Feynman rules
of MSSM vertices with FC effects had to  be implemented  in 
{\it FeynArts}, extending the previous 
model file\footnote{The model file is available on request.}.

The partial widths for the decays $H_x \to b\bar s, s\bar b$ 
($H_x = h^0, H^0, A^0$) 
are simple expressions in terms of the form factors given above.
Assuming that the final states $q \bar q'$ and  $q' \bar q$ 
are experimentally not distinguished, 
the final results for the partial widths are got by adding the 
two individual partial widths, yielding
\begin{eqnarray}
 \Gamma (H_x \to b\bar s+ s\bar b)&=& 
\frac{2g^2}{16\pi m_{H_x}}\sqrt{[1-(\frac{m_s}{m_{H_x}}+\frac{m_b}{m_{H_x}})^2] 
[1-(\frac{m_s}{m_{H_x}}-\frac{m_b}{m_{H_x}})^2]} \nn \\
&&\left[3(m_{H_x}^2-m_s^2-m_b^2)(F_L^{(x)}F_L^{(x) *}+F_R^{(x)}F_R^{(x) *}) \right. \nn\\[0.2cm]
&&\left.-6m_s m_b (F_L^{(x)} F_R^{(x) *}+F_R^{(x)} F_L^{(x) *})\right]\,.
\end{eqnarray}

The dependence of the decay rates and branching ratios on the  MSSM parameters will be 
discussed  in the next section, and 
the behaviour in the large SUSY mass limit 
is investigated analytically and numerically  
in section~\ref{chap.decoupling}.

\section{Numerical analysis}
\label{chap.phenomenology}

Here we numerically estimate the size of the 
loop-induced FCHD as a function of the MSSM parameters and 
the mixing parameter $\lambda$. The GUT relations
$M_3 = \alpha_s/\alpha \, s_W^2\, M_2$ and $M_1=5/3 \, s_W^2/c_W^2 \, M_2$
are assumed. For the numerical analysis of the FCHD rates, 
only values of $\lambda$ (in the range $0\leq \lambda \leq 1$)
that lead to physical squark masses above 150 GeV 
will be considered. The present experimental lower mass bounds on the 
squark masses of the first and second  squark generation
are actually even more stringent that this value~\cite{pdg2002}, 
but we have chosen here this common value of 150 GeV for 
simplicity and definiteness. Notice that higher values of $\lambda$ 
(or, equivalently of $(\delta_{23}^d)_{LL}$ in the usual notation of 
the mass--insertion approximation) are disfavored by 
the present B meson data involving $b$ to $s$ transitions, 
although are not definitely excluded~\cite{9604387,newisidori,ciuchini,Endo}.
Similarly, in view of the present experimental
lower bounds on the chargino mass~\cite{pdg2002}, we will consider only 
$|\mu|$ values above 90 GeV. A lower limit of $M_2 > 54.8$ GeV at $95
\%$ CL, when the chargino neutralino and scalar lepton searches are 
combined, is also considered~\cite{limitM2}.   

The MSSM parameters needed to determine the partial widths 
$\Gamma (H_x \to b \bar s + s \bar b)$, for $H_x \equiv h^0, H^0, A^0$, 
are the following six quantities, 
$m_A$, $\tan \beta$, $\mu$, $M_{2}$, $M_0$, and $A$, where we have 
chosen, for simplicity, $M_0$ as a common value for the soft SUSY-breaking
squark mass parameters, 
$M_0 = M_{\tilde Q,q} = M_{\tilde U,(c,t)} = M_{\tilde D,(s,b)}$,
and all the various trilinear parameters to be universal, $A_t=A_b=A_c=A_s=A$. 
 These parameters will be varied over a broad range, subject only to our 
 requirements that all the squark masses be heavier than 150 GeV,
$|\mu| > 90$ GeV and $M_2 > 54.8$ GeV. In addition, the extra 
parameter $\lambda$ measuring  the FC strength  
will be varied in the range $0 \leq \lambda \leq 1$, by taking into
account the constraints on the squark masses. 
The masses and total decay widths of the Higgs-bosons  have been
computed using the {\it FeynHiggs} program version 
{\it 2.1beta}~\cite{Feynhiggs}.

\begin{figure}[tbh]
\hspace*{-1.6cm}
\begin{center}
\epsfig{file=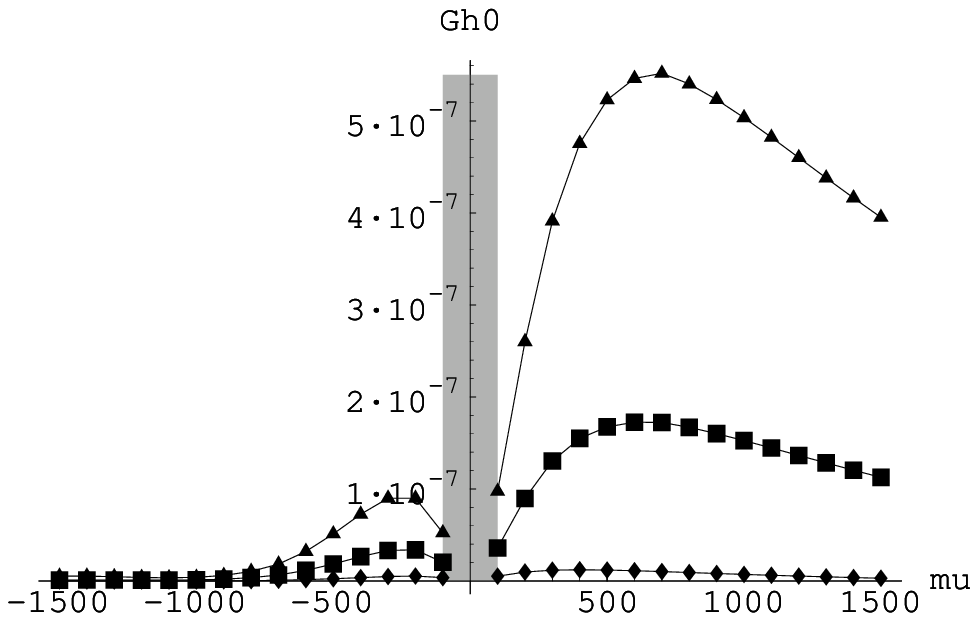,height=2.00in,width=8cm}
\epsfig{file=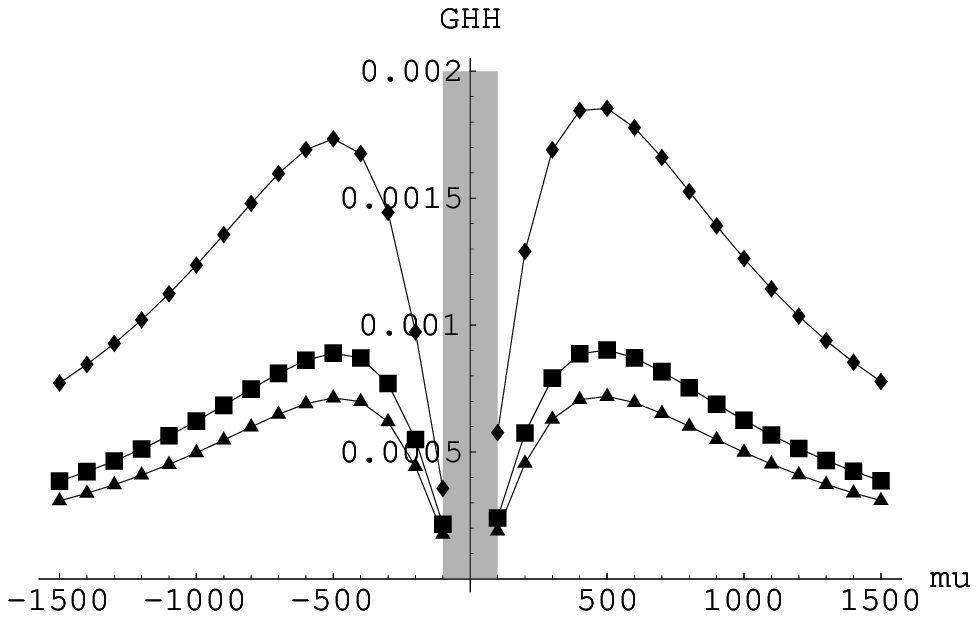,height=2.00in,width=8cm}
\caption{$\Gamma (H_x \to b \bar s + s \bar b)$ $(H_x = h^0, H^0)$
in GeV as a function of $\mu$ (GeV).
Lines with triangles, boxes, and diamonds correspond 
to $m_A=200$ GeV, $m_A=250$ GeV, and $m_A=400$ GeV respectively.}
\label{hbs_mu}
\end{center}
\end{figure}

Figs.~\ref{hbs_1} through~\ref{hbs_M0} display the numerical results for  
$\Gamma (H_x \to b \bar s+ s \bar b)$ as functions of the
MSSM parameters, for the specific value $\lambda=0.5$.
The following default values of the various MSSM parameters, 
\bea
\label{eq:numparameters}
\mu &=& 800 \,\textrm{GeV}\,, M_0=800 \,\textrm{GeV}\,,  A = 500 \,\textrm{GeV} \,,\nonumber\\
m_A &=& 400 \,\textrm{GeV}\,, M_2=300 \,\textrm{GeV}\,, \tan \beta = 35 \, ,
\eea
have been chosen for the figures, 
to specify those parameters that are not varied in each plot.
This set of MSSM parameters is in accordance with
experimental bounds for the decay $b\rightarrow s
\gamma$~\cite{boundsbsg}, as we checked  
with the help of the code {\it micrOMEGAs}~\cite{Omega}, based on
leading order calculations~\cite{leadingbsg} and some
contributions beyond leading order that are important for high
values of $\tan \beta$~\cite{gambinobsg}. 
Nevertheless, for illustration of
interesting dependences, we explore in the following
also a wider range of the MSSM parameters for the decay widths.
\psfrag{MA0}{$\,m_A$}
\psfrag{tb}{{$\tan \beta$}}
\begin{center}
\begin{figure}[t]
\begin{center}
\hspace{-1.6cm}
\epsfig{file=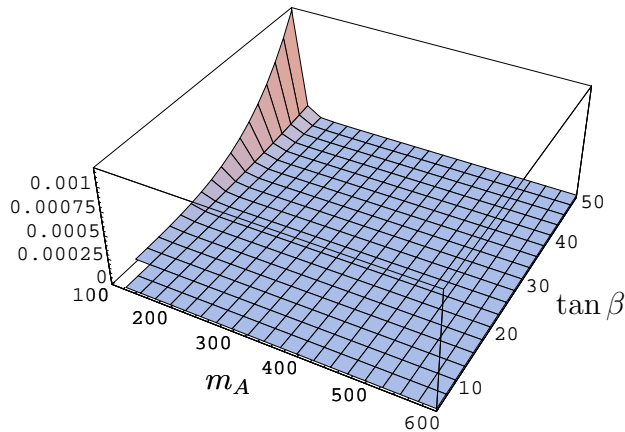,height=2.25in}
\caption{$\Gamma (h^0 \to b \bar s + s \bar b)$ in GeV as a function
of ($\tan \beta$, $m_A$ (GeV)).}
\label{hbs_1}
\end{center}
\end{figure}
\begin{figure}
\hspace{-0.5cm}
\epsfig{file=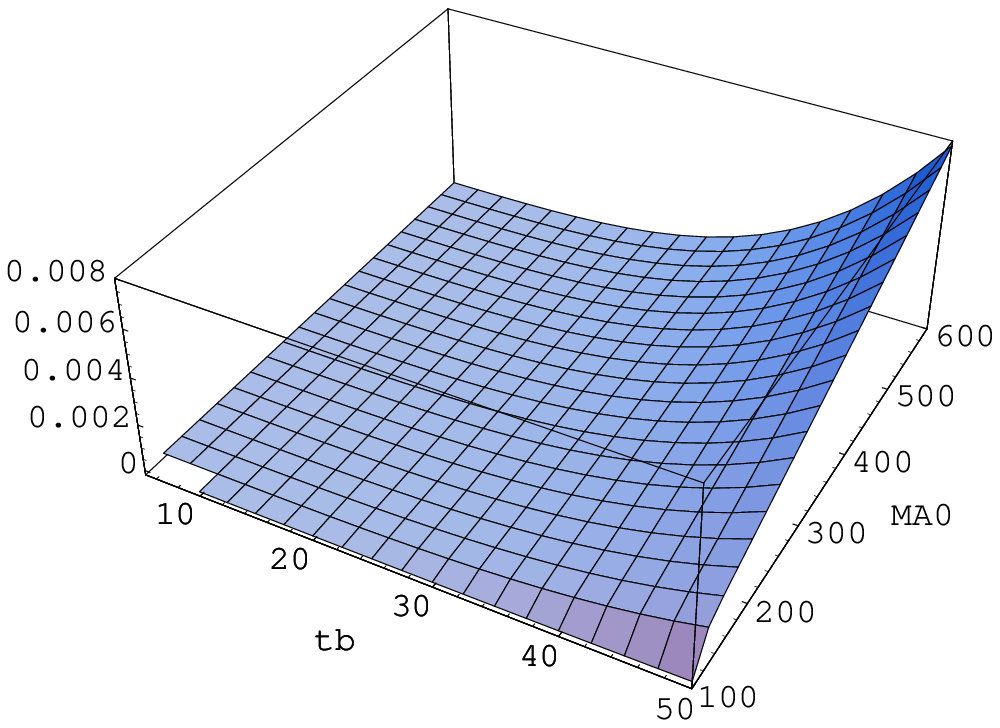,height=2.20in}
\epsfig{file=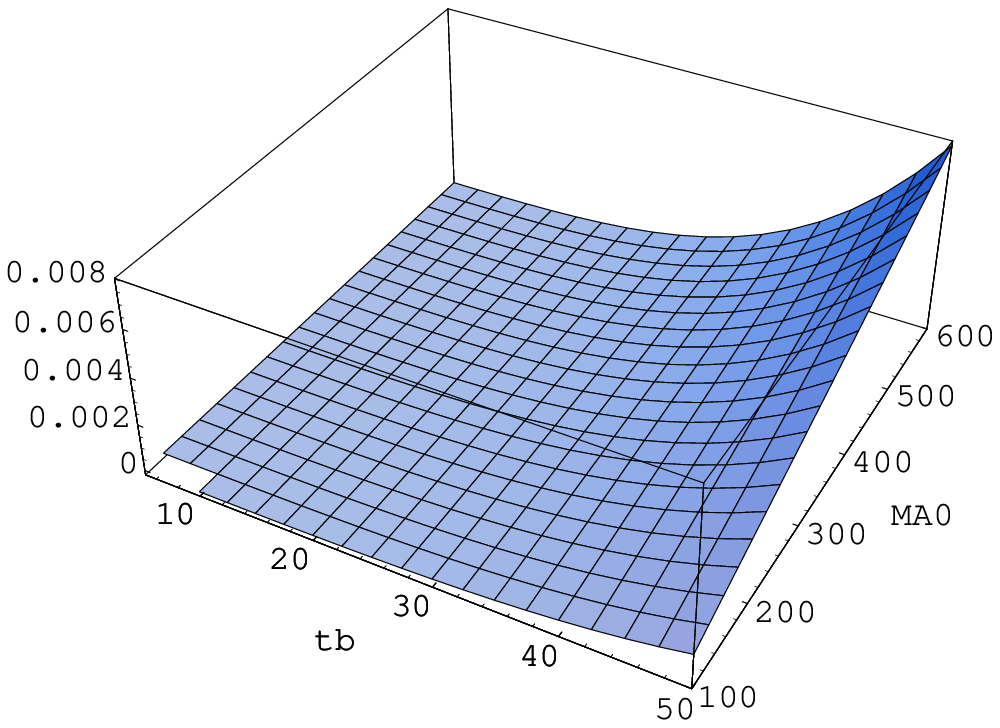,height=2.20in}
\caption{$\Gamma (H^0 \to b \bar s + s \bar b)$ and 
$\Gamma (A^0 \to b \bar s + s \bar b)$ in GeV as a function
of ($\tan \beta$, $m_A$ (GeV)).}
\label{hbs_A0H0}
\end{figure}\vspace*{-0.7cm}
\end{center}

In Figs.~\ref{hbs_1} through~\ref{hbs_M2A}, the MSSM parameters 
have been grouped into pairs ($\tan \beta$, $m_A$) and ($M_2$, $A$) 
in order to visualize the individual dependences
of the FCHD widths for each neutral Higgs boson. 
Figs.~\ref{hbs_1} and ~\ref{hbs_A0H0} show 
$\Gamma (h^0 \to b \bar s+ s \bar b)$, $\Gamma (H^0 \to b \bar s+ s \bar
b)$ and $\Gamma (A^0 \to b \bar s+ s \bar b)$ 
as functions of the pair ($\tan \beta$, $m_A$).
A common  clear behaviour of all three decay widths is the increase 
with $\tan \beta$, yielding maximal FC effects
at large $\tan \beta$ values. In the rest of the
numerical analysis we have chosen $\tan \beta = 35$, as  
specified in (\ref{eq:numparameters}). Notice that the behaviour for 
the $A^0$ decay is indistinguishable from the $H^0$ case. 

The behaviour with $m_A$ is less uniform.  
As one can see from Figs.~\ref{hbs_1} and~\ref{hbs_A0H0},
the decay widths $\Gamma (H^0 \to b \bar s + s \bar b)$ 
and $\Gamma (A^0 \to b \bar s + s \bar b)$ clearly increase with 
$m_A$ owing to obvious phase space effects, i.e. increasing phase space 
for larger Higgs-boson masses. In contrast,  
 $\Gamma (h^0 \to b \bar s + s \bar b)$  shows a less obvious dependence 
on $m_A$. For small values of $m_A$ and large $\tan \beta$ 
we see a sharp decrease of the decay width with $m_A$. 
This is due to the fact that $|\sin \alpha|$ decreases rapidly between 
$m_A = 100$ and 130 GeV and this decrease strongly affects the 
self-energy--like diagrams, which are proportional to 
$\sin \alpha$ (see analytical formulas).
Since the mass of the lightest Higgs boson $h^0$ 
reaches a constant value and the $m_A$-dependence of $\sin \alpha$ is weak
for large values of $m_A$, the decay width is very flat in this range.

\psfrag{M2}{$\,M_2$}
\psfrag{A}{{$A$}}
\begin{center}
\begin{figure}[t]
\hspace{-0.7cm}
\epsfig{file=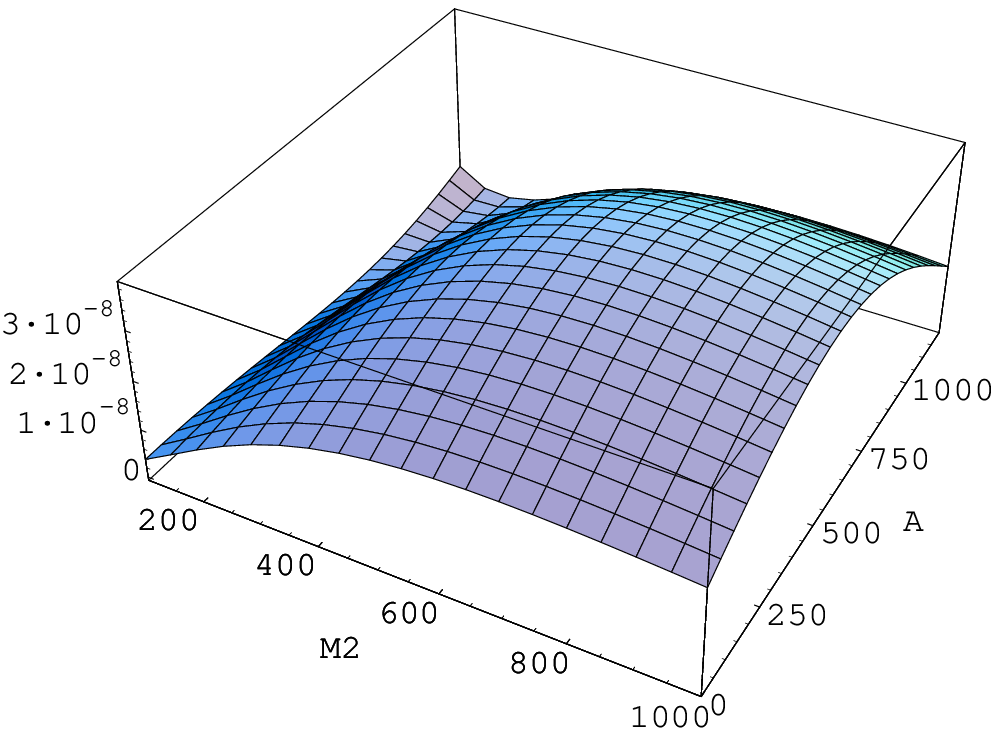,height=2.20in}
\epsfig{file=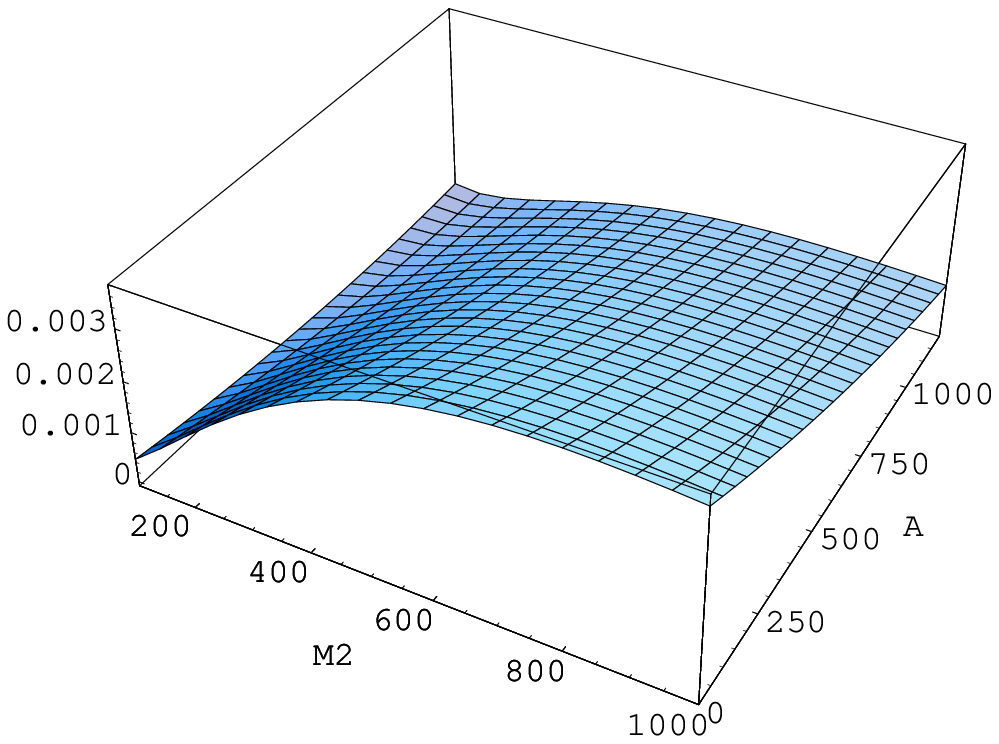,height=2.20in}
\caption{$\Gamma (h^0 \to b \bar s + s \bar b)$  and 
$\Gamma (H^0 \to b \bar s + s \bar b)$ in GeV as a function
of ($M_2$ (GeV), $A$ (GeV)).}
\label{hbs_M2A}
\end{figure}\vspace*{-1.3cm}
\end{center}

Fig.~\ref{hbs_M2A} shows the behaviour of the decay widths with respect to 
($M_2$, $A$). Since the
behaviour for the $A^0$ decay is indistinguishable from the $H^0$ case,
we do not show the $A^0$ decay dependence from now on. Results for the 
$H^0$ decays are applied to this case directly. 
Clearly, for the heavy Higgs boson, the decay width grows with 
$M_2$ up to an approximately constant value and decreases with the
trilinear parameter $A$. 
A less obvious behaviour appears for the $h^0$ case, depending on values
of $M_2$, $A$ and the other MSSM parameters. In this case, we found a similar 
behaviour with respect to $M_2$, but in contrast to the previous case 
we now have an increase of the decay width with growing $A$ up to a
maximum value and then a decrease with this parameter.

Now we focus on the behaviour with respect to the other MSSM
parameters. First, in Fig.~\ref{hbs_mu} we show the 
behaviour of the flavour-changing Higgs decays
$\Gamma (H_x \to b \bar s + s \bar b)$ $(H_x=h^0\,,H^0)$ as functions 
of the $\mu$ parameter for three different values of $m_A$. 
The shaded regions in these figures correspond to the region excluded
by LEP bounds on the chargino mass $|\mu| \lesssim 90$ GeV.
We can see that the width for the $H^0$ decay is approximately symmetric 
under $\mu \to -\mu$, depending of the $m_A$ values. 
In contrast, the $\Gamma (h^0 \to b \bar s + s \bar b)$ width 
is more unsymmetric with respect to the sign of $\mu$.
Note that all decay widths increase with $|\mu|$ for $|\mu| \lesssim 500$ GeV,
then reach a maximum, and finally decrease. Regarding the behaviour at
very small $\mu$ values, we have also found that the widths do not
vanish at $\mu=0$. The origin of this comes entirely from contributions 
driven by electroweak gauge couplings.

\psfrag{GA0}{$\Gamma (A^0 \rightarrow b \bar s + s \bar b)$}
\psfrag{GHH}{$\Gamma (H^0 \rightarrow b \bar s + s \bar b)$}
\psfrag{Gh0}{$\Gamma (h^0 \rightarrow b \bar s + s \bar b)$}
\psfrag{mu}{$\,\mu$}
\psfrag{M0}{$\,M_0$}
\begin{center}
\begin{figure}[t]
\vspace*{0.3cm}
\hspace{-0.9cm}
\begin{center}
\epsfig{file=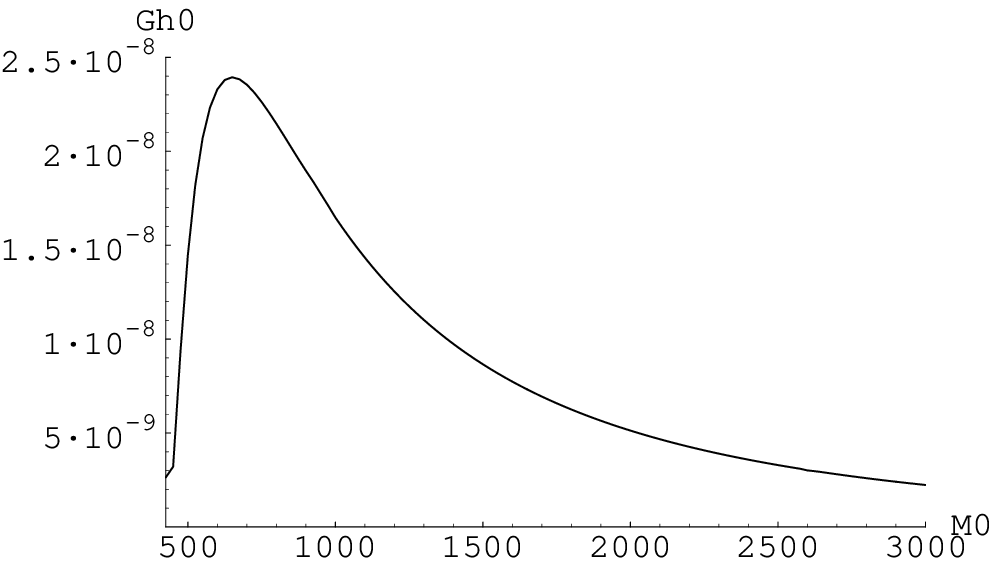,height=5.0cm,width=7.5cm}\hspace*{0.5cm}
\epsfig{file=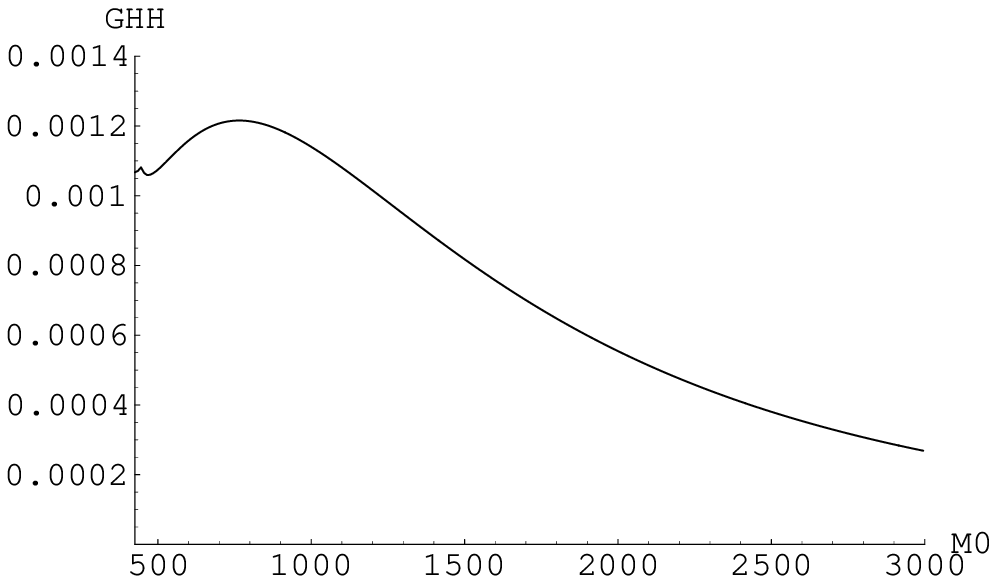,height=5.0cm,width=7.5cm}
\caption{$\Gamma (H_x \to b \bar s + s \bar b)$ as a function of $M_0$ (GeV). 
Other parameters as in~(\ref{eq:numparameters}).}
\label{hbs_M0}
\end{center}
\end{figure}
\vspace*{-0.8cm}
\end{center}

Fig.~\ref{hbs_M0} shows the behaviour of the $h^0$(left panel) and $H^0$ 
(right panel) decays as functions of the common soft SUSY-breaking squark-mass 
parameter $M_0$. The region below $M_0 = 425$ GeV
implies too low and hence forbidden values for the squark masses. 
The $h^0$ decay width has a small value for light $M_0$ due to
the fact that chargino and neutralino contributions have opposite sign
there. For higher values of $M_0$, the neutralino contributions change
sign and the partial cancellation disappears, therefore, the decay width 
increases until it reaches a maximum and then decreases for heavier squarks.
The previously mentioned cancellation for small values of $M_{0}$ is
less obvious for the heavy Higgs, the clearly visible effect is the
decrease due to the growing squark masses. Notice that the decreasing 
behaviour is slower in this case. 

\begin{figure}[t]
\begin{center}\hspace*{-1.0cm}
\includegraphics[height=4.6cm,width=7.5cm,clip=]{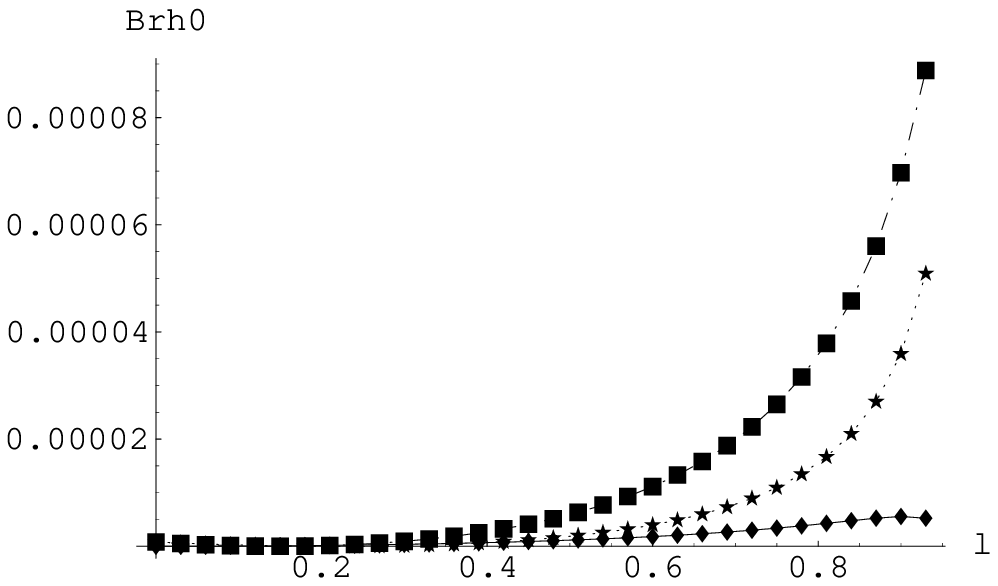}
\includegraphics[height=4.6cm,width=7.5cm,clip=]{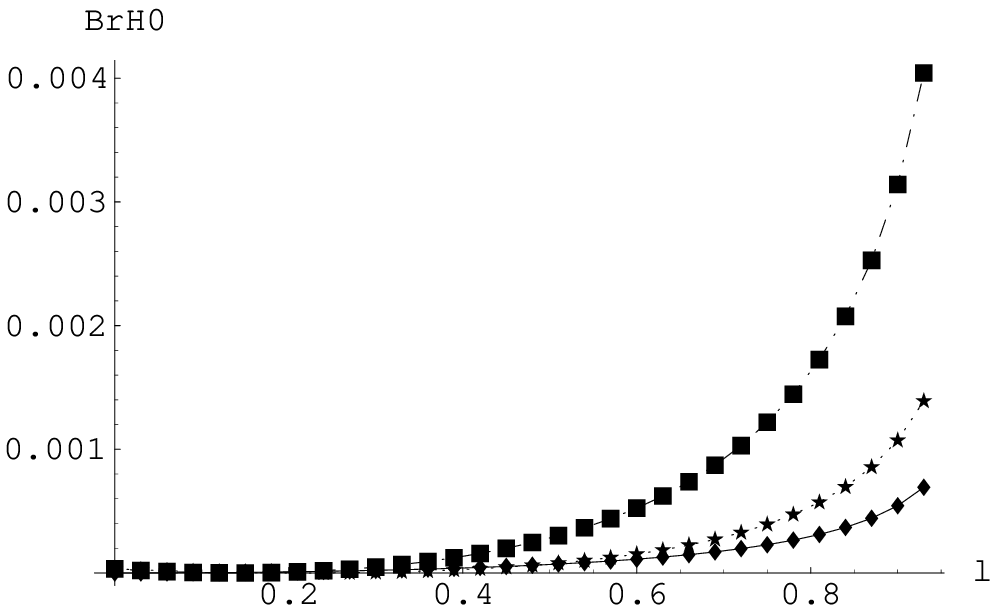}
\caption{$Br(h^0,H^0 \to b \bar s + s \bar b)$  versus 
         $\lambda$ for the parameter set~(\ref{eq:numparameters}). 
         Diamonds (stars) denote neutralino (chargino) contributions, 
         boxes the total SUSY-EW contributions.}
\label{br_hbs_1_all_lambda}
\end{center}
\end{figure}

\begin{figure}[tb]
\begin{center}\hspace*{-1.1cm}
\includegraphics[width=7.5cm,height=4.6cm,clip=]{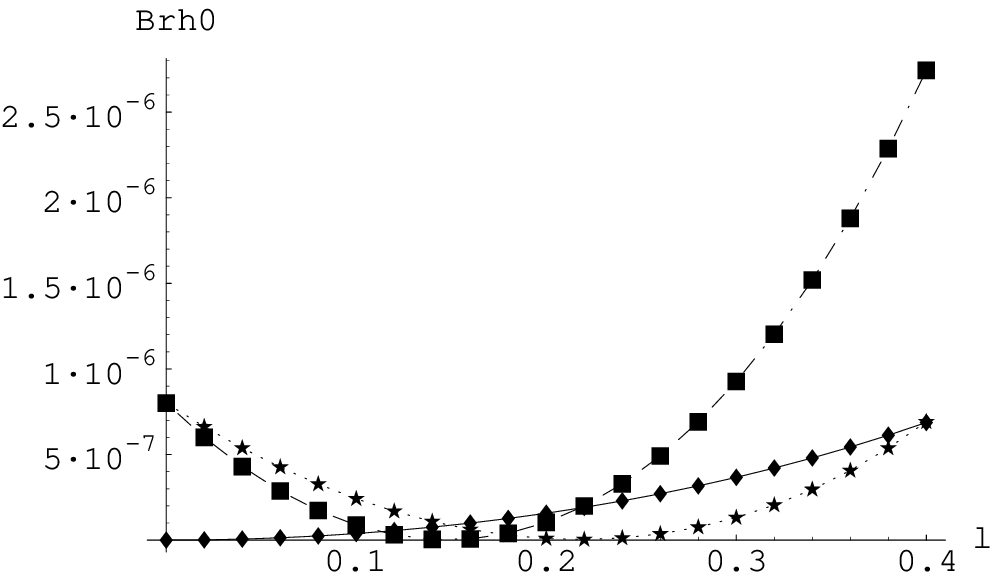}
\includegraphics[width=7.5cm,height=4.6cm,clip=]{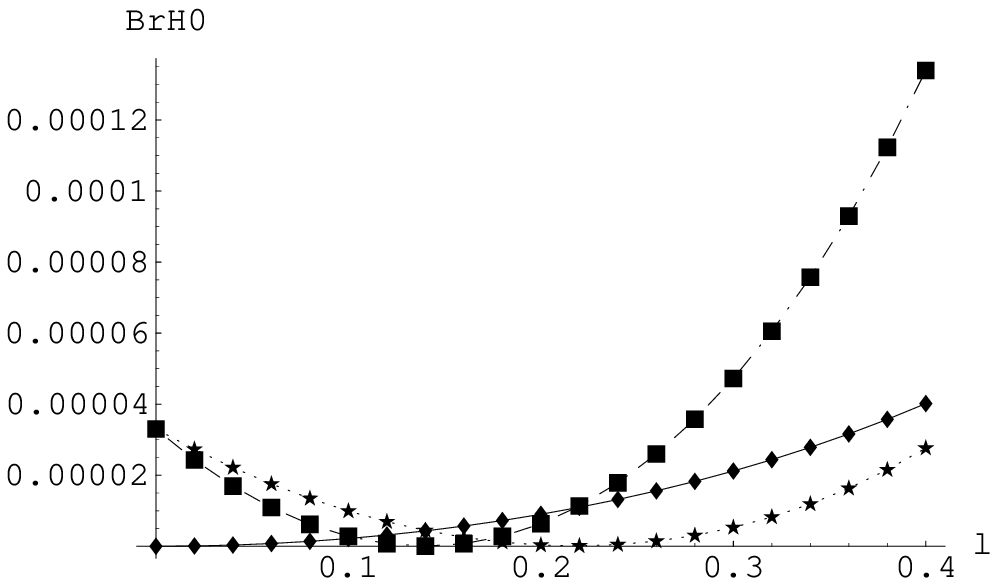}
\caption{Same as in Fig.~\ref{br_hbs_1_all_lambda}, for $0< \lambda< 0.4$.}
\label{br_hbs_1}
\end{center}
\end{figure}

In the following we study the behaviour of the corresponding FCHD with 
respect to $\lambda$. The MSSM parameters are again the ones chosen 
in~(\ref{eq:numparameters}). Given this set of parameters, 
the experimental lower bound on the squark masses restricts
the allowed range of the mixing parameter to $0 \le \lambda \le 0.93$. 
Figs.~\ref{br_hbs_1_all_lambda} and~\ref{br_hbs_1}
contain the branching ratios and show the behaviour individually for the 
neutralino contributions (lines with diamonds), the
chargino contributions (lines with stars), and the total SUSY-EW 
contributions (lines with boxes) up to the maximal allowed value of $\lambda$.
For about $\lambda=0.8$, the branching ratio is around $3\times10^{-5}$ 
for the lightest Higgs boson and $1.7 \times10^{-3}$ for $H^0$ and
$A^0$. We remark that the SM value for this branching ratio is several 
orders of magnitude smaller, yielding 
$Br(H_{SM} \to b \bar s + s \bar b) \sim 4\times 10^{-8}$ for 
$m_{H_{SM}}= 114$ GeV. A similar result can be extracted from~\cite{brinSM}. 

\psfrag{l}{$\lambda$}
\psfrag{Brh0}{$Br (h^0 \rightarrow b \bar s + s \bar b)$}
\psfrag{BrH0}{$Br (H^0 \rightarrow b \bar s + s \bar b)$}

Investigating the contributions from charginos and neutralinos separately, 
we found that the neutralino contribution increases monotonically
with $\lambda$, being exactly zero for $\lambda=0$, as expected. 
On the other hand, the contributions from charginos show explicitly the two FC 
sources, CKM and quark-squark misalignment. In fact,
the chargino contribution is different from zero for $\lambda=0$. 
The non-zero value at $\lambda=0$ is due to CKM mixing which is not present
in neutralino (or gluino) loops.  
The CKM effect and the effect of squark mixing 
for small $\lambda \ne 0$ partially cancel each other, leading to a 
minimum around $\lambda = 0.2$. This minimum depends strongly on 
the particular input for the SUSY mass parameters.  
For larger values of $\lambda$ the non-CKM flavour-mixing effect
dominates and the branching ratio increases,
due to a change of sign in $F_L^{(x)}$ after the minimum.
For our choice for the SUSY mass parameters, eq.~(\ref{eq:numparameters}), 
the total  neutralino contribution to 
the form factors, $F_L$ and $F_R$, is negative for all the studied
values of $\lambda$, but for the chargino contribution  
$F_R$ is always positive and $F_L$ changes sign, being positive for
values of $\lambda \leq 0.2$ and negative for the rest of the studied 
$\lambda$ values. To illustrate this more explicitly, 
we show in Fig.\ref{br_hbs_1} the branching ratios as function of $\lambda$
for the range of small $\lambda$ below 0.4.
The constructive interference of both neutralino and chargino terms 
 lead to a multiple enhancement of the individual contributions in the
 decay rates.

We notice also that $|F_L|$ is larger than $|F_R|$ for both, chargino and 
neutralino contributions. It is also expected since the dominant FC effect 
enters  through the $LL$ entry of the squark mass matrix.  
Regarding  the size of the total chargino/neutralino contributions and which 
diagram is the dominant one, they depend in general on the particular 
Higgs boson and the other MSSM parameters 
and again on the value of $\lambda$. In particular, 
for $\lambda=0.5$, we find that the total chargino and 
 neutralino contributions are of the same order, 
 with the chargino contribution being slightly larger. These
 comments apply to all three Higgs bosons. 
 Finally, for $0.15 <\lambda < 0.4$ the total contribution  
  from neutralinos dominates  that from charginos, also for the three
  Higgs bosons (see Fig.~\ref{br_hbs_1}). 
However, the behavior at very small $\lambda$ values is opposite to the
  previous one, and it is in agreement with the results from 
 the effective Lagrangian approach which incorporates the mass insertion
 approximation~\cite{Demir}. 

\begin{figure}[tb]
\begin{center}\hspace*{-1.1cm}
\includegraphics[width=7.5cm,height=4.6cm,clip=]{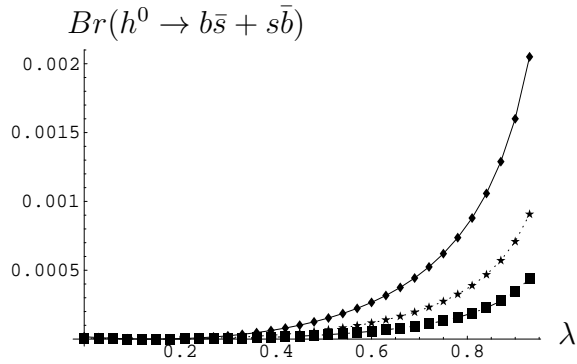}
\caption{$Br ( h^0 \to b \bar s + s \bar b)$ as a function of $\lambda$
  for $m_A=150$ GeV (line with diamonds), $m_A=200$ GeV (line with
  stars) and $m_A=250$ GeV (line with boxes).}
\label{threemA}
\end{center}\vspace*{-0.4cm}
\end{figure}
The results in Fig.~\ref{br_hbs_1_all_lambda} and Fig.~\ref{br_hbs_1}
are for $m_A=400$ GeV. 
We want to emphasize that the decay rates of $h^0$ are 
much larger for smaller values of $m_A$ (see Fig.~\ref{hbs_1}) 
and therefore yield larger values of the branching ratio 
$Br ( h^0 \to b \bar s + s \bar b)$, e.g. 
$Br ( h^0 \to b \bar s + s \bar b) \simeq 2 \times 10^{-3}$ for
$m_A=150$ GeV and $\lambda=0.85$. 
To illustrate the $m_A$ dependence, we show in Fig.~\ref{threemA} 
the results for this branching ratio as a
function of $\lambda$ for $m_A=150\,, 200\,, 250$ GeV. 
Large branching ratios are found for small $m_A$ and large $\lambda$ values. 

\begin{figure}[t]
\begin{center}
\hspace*{-1.0cm}
\epsfig{file=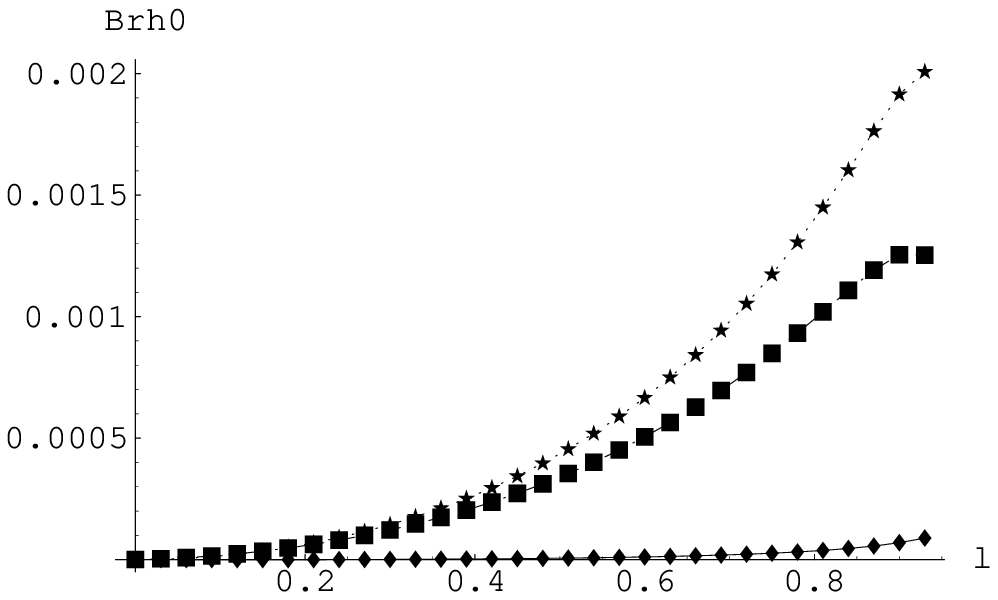,height=4.5cm,width=7.6cm} 
\epsfig{file=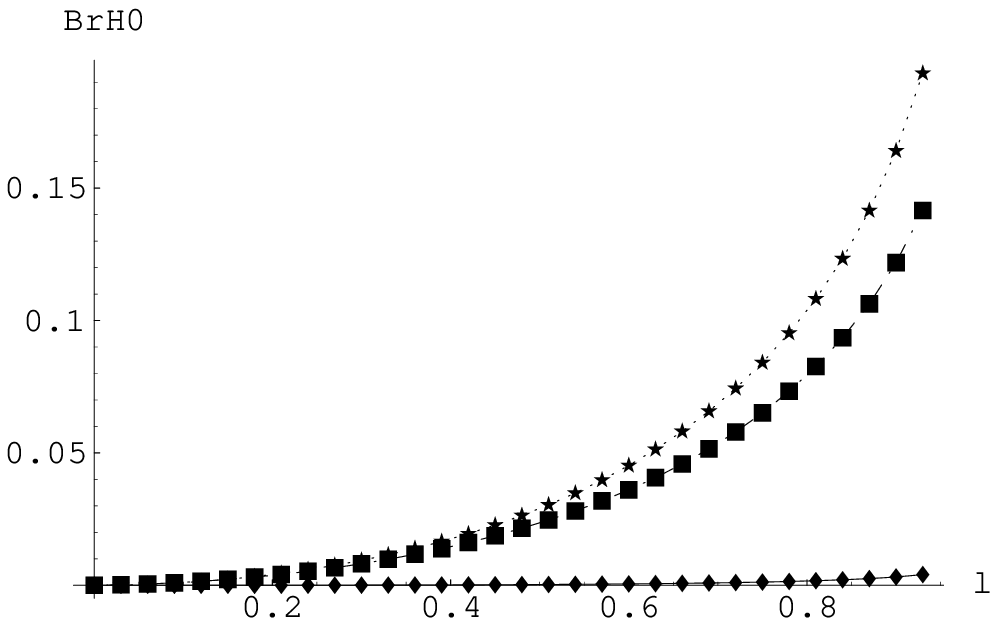,height=4.5cm,width=7.6cm}
\caption{SUSY-EW (Diamonds), SUSY-QCD (stars) and total (boxes) contributions
  to the $Br( H_x \to b \bar s + s \bar b)$ $H_x =(h^0, H^0)$  as a 
function of $\lambda$ for the selected MSSM parameters.}
\label{ewglu}
\end{center}
\end{figure}

In summary, the  FCHD branching ratios that we have found in this section 
are quite sizable, and are in fact some orders of magnitude larger than the 
corresponding SM rates, but small in comparison with 
the SUSY-QCD contributions computed previously in~\cite{Maria}.
There are, however, important interference terms, which modify the
SUSY-QCD effects remarkably. The combined results for the SUSY-EW, SUSY-QCD 
and the total contributions
to the $Br( H_x \to b \bar s + s \bar b)$ $H_x =(h^0, H^0)$ are
displayed as a function of $\lambda$
in Fig.~\ref{ewglu}. 
 Note that the absolute value for 
gluino, chargino (above the minimum) and neutralino contributions grow with 
$\lambda$ separately.
The SUSY-QCD quantum contributions are at least one order
of magnitude larger than the pure SUSY-EW contributions. However, these later 
contribute with opposite sign, providing an important interference effect.

\section{Non-decoupling behaviour of heavy SUSY particles}
\label{chap.decoupling}

In this section we study the non-decoupling behaviour of squarks, 
charginos and neutralinos of the SUSY-EW contributions to FCHD
  into $b \bar s$ and $s \bar b$. This non-decoupling behaviour of
 the SUSY particles also happens in the SUSY-QCD
 contributions~\cite{Maria} and means that the FC effects remain 
non-vanishing even in the most pessimistic scenario of a very heavy 
SUSY spectrum. The motivation to analyze these effects, including both 
the SUSY-EW and the SUSY-QCD contributions, is that they could provide 
a very efficient way to search for indirect SUSY signals at next 
generation colliders. 

The origin of this non-decoupling behaviour in the SUSY-EW contributions
is, as in the SUSY-QCD case, the fact that the mass suppression induced 
by the heavy-particle propagators is compensated by the mass parameter 
factors coming from the interaction vertices, this being generic in 
Higgs--boson physics. It has been previously analyzed at length in the case
 of flavour-preserving MSSM Higgs 
decays~\cite{ourHtb,othersHtb,MJsitges,dobado,siannah}.
The non-decoupling contributions to effective FC Higgs Yukawa couplings have 
also been studied in the effective-Lagrangian approach for the quark 
sector~\cite{Demir,dedes} and the leptonic sector~\cite{Brignole}. In this 
approach, the FC effects are encoded in a set of nonholomorphic 
dimension-four effective operators which appear due to SUSY breaking 
induced from the radiative corrections. It usually considers just the 
dominant contributions that come 
from large Yukawa couplings and assumes large $\tan \beta$ values. 
It is simple, but can not easily implement $SU(2) \times U(1)$
electroweak symmetry breaking effects
which are relevant for small values of $\tan \beta$. We will use here instead
the full diagrammatic approach which has the advantage of taking into account
  all EW loop contributions and is valid for all
$\tan \beta$ values. Since, on the other hand,
 we are not using the mass--insertion approximation, our results are more 
general, being valid for all values of the FC parameter $\lambda$. We 
will see that our results converge in the large $\tan \beta$ limit and 
for small $\lambda$ values to the mass insertion approximation results of  
the effective Lagrangian approach.

In order to show analytically the non-decoupling behaviour of squarks, 
charginos and neutralinos in the FCHD, we perform a systematic expansion 
of the form factors involved, and hence
of the partial widths, in inverse powers of the heavy SUSY masses and look for
the first term in this expansion.
We have considered the simplest hypothesis for the SUSY masses where all the 
soft breaking squark mass parameters, collectively denoted by $M_{0}$, the 
$\mu$ parameter, the trilinear parameters, collectively denoted by $A$ and 
the gaugino masses, are chosen to be of the same order and much greater than 
the electroweak scale $M_{EW}$,
\begin{equation}
M_{S} \sim M_{0} \sim M_{\tilde g} \sim M_1 \sim M_2 \sim \mu 
\sim A  \gg M_{EW},
\label{eq.largeSUSYlimit}
\end{equation}
where $M_{0} = M_{\tilde Q} = M_{\tilde U} = M_{\tilde D}$ 
and $M_{\tilde g}$ is the gluino mass. The bino and wino soft-breaking
 masses, $M_1$ and $M_2$, are chosen in our numerical evaluations to 
follow the GUT relations, 
$M_1 = \gamma M_{\tilde g}$ and $M_2 =\eta M_{\tilde g}$,
 with $\gamma = \frac{5}{3}\frac{g_1^2}{g_3^2}$ and $\eta =\frac{g^2}{g_3^2}$.
In order to provide more compact analytical results that can be easily used 
for future phenomenological studies, we have also analyzed the case of 
equal SUSY mass parameters,
$M_{S} = M_{0} = M_{\tilde g}= M_1 = M_2 =\mu = A$, where  $\gamma=\eta=1$. 

In this section, we present the expansions of the SUSY-EW contributions to 
the form factors in inverse
powers of $M_S$ and keep just the leading contribution of this expansion 
by considering that all the remaining involved mass scales 
$m_{H^0},m_{A},m_{h^0},m_Z,m_W$ and $m_q$  are of order $M_{EW}$. To this end,
we use the results of the expansions of the one-loop functions and of the 
rotation matrices that are given in Appendix B.
In the general case of arbitrary $\gamma$ and $\eta$, we find the
analytical results for the various contributions coming from 
chargino-squark and neutralino-squark
 loops that are collected in Appendix C. We next comment these results.
 First, we can see from eqs.(\ref{formfactorLdecoupequalescalebs1}) 
 to (\ref{formfactorLdecoupequalescalebs7}) that, taking
 all SUSY mass parameters arbitrarily large and of the same
order, ${\cal O}(M_S)$, the SUSY-EW contributions to the form factors 
lead to a non-zero value. That is, they do not decouple in
the large SUSY mass scenario. Regarding the ultraviolet behaviour,
  we have checked that the various contributions shown in formulas 
 (\ref{formfactorLdecoupequalescalebs1}) 
 to (\ref{formfactorLdecoupequalescalebs7}) are separately finite. 
A discussion on the relative signs and the sizes of chargino and
neutralino contributions to the form factors, $F_L$ and $F_R$, is
already included in the previous section. In general, the approximate 
analytical results show similar features than the exact results.

\begin{figure}[t]
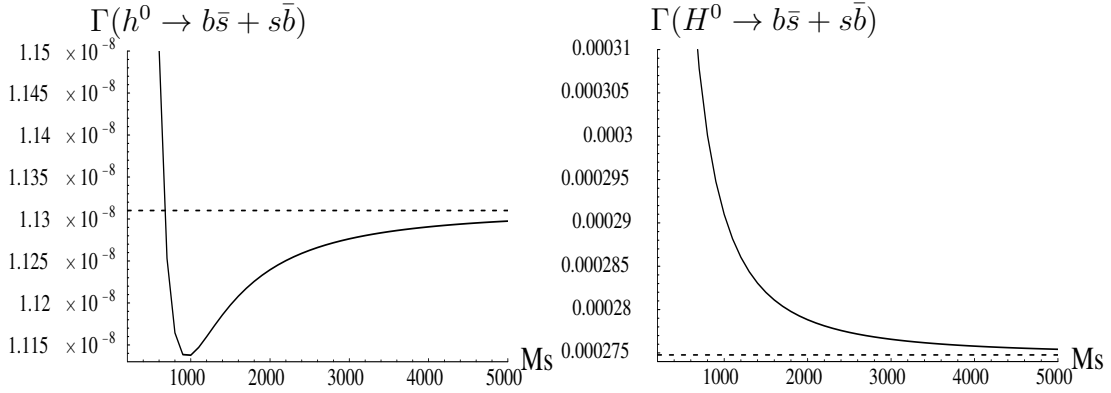

\hspace{-1.0cm}
\begin{center}
\epsfig{file=fig11a.epsi,height=5.0cm,width=7.2cm}
\epsfig{file=fig11b.epsi,height=5.0cm,width=7.2cm}
\caption{Non-decoupling behaviour of $\Gamma(H_x \to b\bar s + s\bar b)$ in GeV with 
$M_0=\mu=A=M_{\tilde g}=M_S$, $M_1 = \frac{5}{3}\frac{g_1^2}{g_3^2}  M_{\tilde g}$ 
and $M_2 =\frac{g^2}{g_3^2}  M_{\tilde g}$,
for $H_x = h^0$ (left panel) and $H_x =H^0$ (right panel) 
and for $\tan\beta=35$, $\lambda = 0.5$, $m_{A}=m_{H^0}=250$ GeV and
$m_{h^0}=135$ GeV. Exact one-loop results in solid lines and expansions
given in eq. (\ref{formfactorLdecoupequalescalebs1}) in dashed lines 
are plotted for comparison.}
\label{widthdecouphbs_Ms}
\end{center}
\end{figure}

In the following we show graphically some of these features.   
For definiteness, in all these plots we choose  
 $M_0 = M_{\tilde g} = \mu = A=M_S$, $M_1 = 
\frac{5}{3}\frac{g_1^2}{g_3^2} M_{S}$, 
 $M_2 =\frac{g^2}{g_3^2} M_{S}$, $\tan\beta=35$ and,
in order to simplify the analysis, we fix the Higgs boson masses to the 
following particular values, $m_{h^0} = 135$ GeV, $m_{H^0} = 250$ GeV 
and $m_{A} = 250$ GeV.  The non-decoupling behaviour 
of $\Gamma(H_x \to b\bar s + s\bar b)$, for $H_x = h^0$ (left panel) and 
$H_x =H^0$ (right panel), as a function of $M_S$ is illustrated in 
Fig.~\ref{widthdecouphbs_Ms}. Here we have fixed  $\lambda=0.5$.
 The exact one-loop results (solid lines) and the approximate large $M_S$ 
 analytical expansions of eqs.~(\ref{formfactorLdecoupequalescalebs1}) 
to~(\ref{formfactorLdecoupequalescalebs7})  (dashed
 lines) are shown for comparison. The behaviour of 
$\Gamma(A^0 \to b\bar s + s\bar b)$ is practically identical to that of 
$\Gamma(H^0 \to b\bar s + s\bar b)$ and is not shown for brevity. 
 We can see in these figures that, for large values of $M_S$, the exact partial
  widths tend to a non-vanishing value, characteristic of the non-decoupling
  behaviour,  which is very well described by our asymptotic results. 
 Notice that a proper study of the non-decoupling behaviour requires to
take the input Higgs boson parameters, i.e the masses and the $\alpha$
mixing angle, at their tree level values. However, for illustrative
purposes in the case of the lightest Higgs boson, we have prefered not to
use the less phenomenologically appealing tree level value but to take
instead an effective input mass value of 135 GeV which includes roughly
the leading one-loop corrections for $\tan \beta=35\,, m_{A^{0}} = 250$
GeV and $M_S \sim O(1 TeV)$.
Notice also that $M_S$ has been taken  up to very 
large values just to illustrate the non-decoupling behaviour, but the  
  convergence of the exact result to the asymptotic one is very good  
 already at moderate $M_{SUSY}$ values (say $\geq 600$ GeV), what 
makes our asymptotic formulas useful for future 
phenomenological Higgs boson studies.   

\begin{figure}[t]
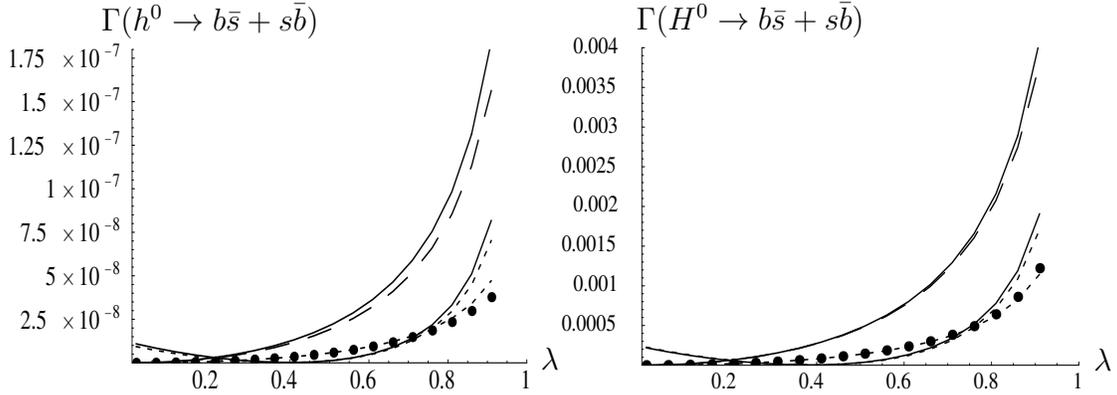

\hspace{-1.0cm}
\begin{center}
\epsfig{file=fig12a.epsi,height=5.0cm,width=7.2cm}
\epsfig{file=fig12b.epsi,height=5.0cm,width=7.2cm}
\caption{$\Gamma (H_x \to b \bar s + s \bar b)$
  as a function of $\lambda$. Left panel shows the $h^0$ decay and right panel
corresponds with the $H^0$ decay. Exact one-loop chargino contribution in 
solid lines. The largest contribution corresponds to the effect coming 
only from misalignment; in the other, both possible effects are included. 
The neutralino contribution in dots. 
The expansions given in eq.~(\ref{formfactorLdecoupequalescalebs1} 
to~\ref{formfactorLdecoupequalescalebs7}) in dashed lines 
are plotted for comparison. The fixed parameters are taken as in 
Fig.~\ref{widthdecouphbs_Ms}.}\vspace*{-0.3cm}
\label{widthdecouphbs_lambda}
\end{center}
\end{figure}

The dependence of the total chargino and neutralino contributions    
on the FC parameter $\lambda$ are shown in Fig.~\ref{widthdecouphbs_lambda}.
Here we have fixed $M_S=1000$ GeV. The four lower lines take into account 
the two FC effects, quark mixing from off-diagonal terms in the CKM matrix 
and squark mixing from quark--squark misalignment, and show the comparative 
size of neutralino and chargino contributions. The solid (dotted) lower 
line is the total exact chargino (neutralino)
contribution. The dashed lower lines are the corresponding approximate results
given by our asymptotic formulas of Appendix C. First, we see that our
asymptotic formulas describe extremely well the behavior with the $\lambda$
parameter for the whole studied interval, $0\le \lambda \le 1$, and this
is true for both chargino and neutralino contributions. Second, the 
neutralino and chargino contributions behave as expected at very small 
$\lambda$ values. As has been already discussed in
section~\ref{chap.phenomenology}, the
chargino contributions are larger than the neutralino ones at these small $\lambda$ values, and at     
$\lambda =0$  the first one is non-vanishing whereas the second one
vanishes. For larger $\lambda$ values the
situation is different. Here we find that for $ 0.2 \le \lambda \le 0.7$ the neutralino
contribution is larger than the chargino contribution, which has now a minimum at about 0.4, and for $\lambda \ge 0.7$
again the chargino contribution dominates. As has been mentioned in section~\ref{chap.phenomenology}, the localization of this minimum varies with the choice of MSSM parameters. Notice that for moderate and 
large $\lambda$ values the alternative and more frequently used mass insertion
approximation fails. It is clear from our plots that for these  
$\lambda$ values to assume a linear behaviour with 
$\lambda$ in the form factors is not correct and indeed can give a wrong conclusion on selecting 
the dominant contributions.

The two upper lines in Fig.~\ref{widthdecouphbs_lambda} are the chargino
contributions for the simplified case of $V_{CKM}=1$, which means that the only
source of FC effects in this case is quark--squark misalignment.  We see
that the contribution vanishes at $\lambda =0$, as expected.  Again the solid
line is the exact result and the dashed line is the approximate asymptotic
result. As in the previous case, we see the excellent agreement of our large 
$M_S$ results with the exact ones for all $\lambda$ values. The interesting
feature here is to compare these lines with the previous ones and to notice 
how large is the effect of misalignment as compared to CKM in the chargino
contributions. Indeed, we see that, for $\lambda \ge 0.2$, these two 
effects enter with different signs
and give a total chargino contribution that is considerably lower than if we
switch off the effects from CKM. This reduction of the chargino contribution
is what makes finally the two contributions, from charginos and neutralinos, to
be of comparable size, and therefore to neglect the later is not a good
approximation.   

Finally, we present the results for the case where all SUSY mass
parameters are equal. These are very simple formulas that, besides of 
being useful for future phenomenological studies, allow us to compare 
more easily our results with those of the effective Lagrangian
approach. The separate results for the various contributing diagrams  
are given in eqs.~(\ref{formfactorLdecoupequalescalebs9}) 
to~(\ref{formfactorLdecoupequalescalebs14}) of Appendix C. 
The total chargino and neutralino contributions
to the form factors for equal mass scales are then given by
\begin{eqnarray}
&&\hspace*{-0.5cm}F_{L_{\tilde \chi^{\pm}}}^{(x)}= 
\frac{\alpha_{EW}}{4 \pi} \frac{m_b}{2 m_W \cos\beta}\left[\frac{1}{8 m_W^2 \sin ^2 \beta} \left[\left(V_{{\scriptscriptstyle CKM}}^{tb} V_{{\scriptscriptstyle CKM}}^{cs} m_c^2 +V_{{\scriptscriptstyle CKM}}^{cb} V_{{\scriptscriptstyle CKM}}^{ts} m_t^2 \right)F(\lambda) 
\right.\right. \nn \\
&& \left. 
+ \left(V_{{\scriptscriptstyle CKM}}^{cb} V_{{\scriptscriptstyle CKM}}^{cs} m_c^2 +V_{{\scriptscriptstyle CKM}}^{tb} V_{{\scriptscriptstyle CKM}}^{ts} m_t^2 \right)J(\lambda)\right] -\frac{1}{4} \left[ \left(V_{{\scriptscriptstyle CKM}}^{cb} V_{{\scriptscriptstyle CKM}}^{cs} + V_{{\scriptscriptstyle CKM}}^{tb} V_{{\scriptscriptstyle CKM}}^{ts}\right) J(\lambda)
\right. \nn \\
&& \left. \left. 
+\left(V_{{\scriptscriptstyle CKM}}^{cb} V_{{\scriptscriptstyle CKM}}^{ts} + V_{{\scriptscriptstyle CKM}}^{tb} V_{{\scriptscriptstyle CKM}}^{cs}\right) F(\lambda)\right]\right] \left(\sigma_2^{(x)}+\tan\beta\, \sigma_1^{(x)*} \right)\,,
\label{eq.hbsLLonescalecharginos} \\ \nn \\
&&\hspace*{-0.5cm}F_{L_{\tilde \chi^0}}^{(x)} = 
-\frac{\alpha_{EW}}{4 \pi} \frac{m_b}{2 m_W \cos\beta}
\left[\frac{1}{8}\left(1+\frac{5}{9}\tan^2\theta_W\right)\left(\sigma_2^{(x)}+\tan\beta\, \sigma_1^{(x)*} \right)\right]F(\lambda)\,,
\label{eq.hbsLLonescaleneutralinos} 
\end{eqnarray} 
where $\alpha_{EW}=\frac{g^2}{4 \pi}$. The results for $F_R$ are 
like the previous ones but replacing 
$m_b\rightarrow m_s$, $m_c\rightarrow m_t$ and taking the complex
conjugate. Notice that in eq.~(\ref{eq.hbsLLonescalecharginos}) we can 
see clearly the relative importance of the two FC effects, 
quark--squark misalignment and CKM, which are driven
respectively by the two functions $F(\lambda)$ and $J(\lambda)$ whose 
values at the origin are given by  $F(0)=0$ and $J(0)=2$. We can also
learn from eqs.~(\ref{eq.hbsLLonescalecharginos}) 
and~(\ref{eq.hbsLLonescaleneutralinos}) about the relative size of the 
contributions from Yukawa couplings versus those from pure gauge
couplings. For instance, for $\lambda=0.5$, $m_A=250$ GeV and $\tan
\beta=35$, the contributions from pure gauge couplings are about $50\%$ of the ones from Yukawa couplings and of opposite sign.

On the other hand, if we keep just the contributions from Yukawa
couplings and neglect the contributions from pure gauge
couplings, only the first term in eq.~(\ref{eq.hbsLLonescalecharginos}), 
which goes with the quark masses, remains.  
If we now consider the large $\tan \beta$  limit and the small $\lambda$ limit
of the previous result, take $V_{{\scriptscriptstyle CKM}}=1$, and use
the linear approximation where $F(\lambda)\simeq - \frac{2}{3}\lambda$, we get,
 \begin{eqnarray}
&&\hspace*{-0.5cm}F_L^{(x)} = -
\frac{\alpha_{EW}}{4 \pi} \frac{m_b}{2 m_W \cos \beta}\left( 
\frac{m_c^2}{12 m_W^2} \tan\beta\, \sigma_1^{(x)*}  \right)\lambda \,,
\end{eqnarray}
which agrees with the result from the effective-Lagrangian 
approach~\cite{Demir}.

Finally, we briefly comment on the interesting limit called decoupling limit where $m_A>>m_{EW}$, which corresponds to a Higgs sector with very heavy Higgs bosons except $h^{0}$. Notice that when $x=h^{0}$, the common factor, $(\sigma_2^{(x)}+\tan\beta\, \sigma_1^{(x)*})$ in  eqs.~(\ref{eq.hbsLLonescalecharginos}) and~(\ref{eq.hbsLLonescaleneutralinos}) is $(\cos \alpha + \tan \beta \sin \alpha)$ which in this decoupling limit goes to zero. Therefore, the decoupling of the heavy particles indeed occur in this case and we recover the SM vanishing value, as expected.

\section{ Conclusions}

In this work we have computed the SUSY-EW quantum effects to 
flavour-changing MSSM Higgs-boson decays into $b \bar s$ 
and $s \bar b$. We have used
 the full diagrammatic approach and therefore our results are valid for all
 $\tan \beta$ values and for all values of the 
flavour-mixing parameter $\lambda$. We 
analyzed in full detail the dependence of the FCHD partial widths, with all 
the relevant MSSM parameters and $\lambda$, and found that they are very 
sensitive to $\tan \beta$, $\mu$, and $\lambda$. The branching ratios 
grow with both $\tan \beta$ and $\lambda$ and reach quite sizable values 
in comparison with the SM ones, in the large $\tan \beta$ and $\lambda$ 
region. For instance, for $\tan \beta =35$, $m_A = 400$ GeV and 
$\lambda=0.8$, we found branching ratios of $3 \times 10^{-5}$ for
 the $h^0$ and $1.7 \times 10^{-3}$ for the $A^0$ and $H^0$. These SUSY-EW 
contributions are subdominant with respect to the SUSY-QCD contributions,
but they contribute with opposite sign and important interference
effects, which modify the SUSY-QCD effects remarkably, appear.

The most interesting feature of these SUSY-EW radiative contributions is their 
non-decoupling behaviour for large values of the SUSY particle masses. We have
 analyzed this behaviour in great detail and found that these SUSY-EW 
contributions to the FCHD widths, as the SUSY-QCD ones, indeed
 do not vanish for asymptotically large SUSY mass parameters. We presented 
a set of analytical asymptotic results that can be of great utility for 
future phenomenological studies. These results are in agreement with the 
ones of the effective--Lagrangian approach in the large $\tan \beta$ and 
small $\lambda$ limit. However, we checked that, for moderate $\tan \beta$ 
and $\lambda$ values, some of the contributions which are usually neglected 
in the effective--Lagrangian approach can be sizable and, therefore, a 
realistic estimate of the branching ratios should rely better on the
full diagrammatic approach. In particular, the contributions driven from 
pure gauge couplings should not be ignored for these moderate values. 

In conclusion, the results presented in this work indicate that a 
phenomenological detailed study of the Higgs boson decays into $b \bar s$ 
and $s \bar b$ can be very efficient as an indirect method in the future 
search for supersymmetry.  

\section*{Acknowledgments}

This work was supported in part by the European Community's Human
Potential Programme under contract HPRN-CT-2000-00149 
``Physics at Colliders'' and by the Spanish MCyT under project 
FPA2003-04597. A.M.C acknowledges MECD for financial support by 
FPU Grant No. AP2001-0678 and D. Temes for many useful discussions. 

\section*{Appendix A}
\setcounter{equation}{0}
\renewcommand{\theequation}{A.\arabic{equation}}

Here are the abbreviations used in the analytical expressions for the form factors $F_{L,R}^{(x)}$ (\ref{formfactorLbs}). 
\begin{eqnarray*}
     \kappa_{L 1}^{x,\,\tilde \chi^0} = B_{L \alpha a}^{(b)}D_{R
     ab}^{(x)}E_{L \alpha b}^{(s)} && \iota_{L 1}^{x, \tilde\chi^0} = B_{R \alpha a}^{(b)}E_{L \beta a}^{(s)}\\
     \kappa_{L 2}^{x,\,\tilde \chi^0} = B_{R \alpha a}^{(b)}D_{L
     ab}^{(x)}E_{R \alpha b}^{(s)} && \iota_{L 2}^{x, \tilde\chi^0} = B_{L \alpha a}^{(b)}E_{R \beta a}^{(s)}\\
     \kappa_{L 3}^{x,\,\tilde \chi^0} = B_{R \alpha a}^{(b)}D_{L
     ab}^{(x)}E_{L \alpha b}^{(s)} && \iota_{L 3}^{x, \tilde\chi^0} = B_{L \alpha a}^{(b)}E_{L \beta a}^{(s)}\\
     \kappa_{L 4}^{x,\,\tilde \chi^0} = B_{L \alpha a}^{(b)}D_{R
     ab}^{(x)}E_{R \alpha b}^{(s)} && \\
     \kappa_{L 5}^{x,\,\tilde \chi^0} = B_{R \alpha a}^{(b)}D_{R ab}^{(x)}E_{L \alpha b}^{(s)} &&\\
     \kappa_{L 6}^{x,\,\tilde \chi^0} = B_{L \alpha a}^{(b)}D_{L ab}^{(x)}E_{R \alpha b}^{(s)} &&\\
     \kappa_{L 7}^{x,\,\tilde \chi^0} = B_{L \alpha a}^{(b)}D_{L
     ab}^{(x)}E_{L \alpha b}^{(s)} &&
\end{eqnarray*}\vspace*{-0.7cm}
\begin{eqnarray}
\Sigma_L^{\tilde\chi^0} (p^2) &=&
-\frac{g^2}{16\pi^2}B_1(p^2,m_{\tilde\chi_a^0}^2, m_{\tilde d_\alpha}^2)
       B_{R\,\alpha a}^{(b)}E_{L\,\alpha a}^{(s)} \nn \\
     m_b \Sigma_{Ls}^{\tilde \chi^0} (p^2) &=&  \frac{g^2m_{\tilde\chi_a^0}}{16\pi^2}B_0(p^2,m_{\tilde\chi_a^0}^2, m_{\tilde d_\alpha}^2)
       B_{L\,\alpha a}^{(b)}E_{L\,\alpha a}^{(s)} 
\end{eqnarray}

The $\kappa_R^{x, \, \tilde\chi^0}$,  $\iota_R^{x, \, \tilde\chi^0}$, $\Sigma_R^{\tilde\chi^0}$ can be obtained from   
$\kappa_L^{x, \, \tilde\chi^0}$,  $\iota_L^{x, \, \tilde\chi^0}$,
$\Sigma_L^{\tilde\chi^0}$ respectively by exchanging all indices $(L
\leftrightarrow R)$.

For chargino contributions, the $\kappa^{x, \, \tilde\chi^-}$,
$\iota^{x, \, \tilde\chi^-}$, $\Sigma^{\tilde\chi^-}$  can be
obtained from the previous expressions for
$\kappa^{x, \, \tilde\chi^0}$,  $\iota^{x, \, \tilde\chi^0}$, $\Sigma^{ \tilde\chi^0}$ by making the replacements 
$m_{\tilde\chi_a^0}\rightarrow m_{\tilde\chi_i^-}$, $m_{\tilde d_\alpha}\rightarrow m_{\tilde u_\alpha}$, 
$B^{(b)}\rightarrow A^{(b)}$, $E_L^{(s)}\rightarrow A_R^{(s)*}$, $E_R^{(s)}\rightarrow A_L^{(s)*}$, $D^{(x)}\rightarrow W^{(x)}$, $a \rightarrow i$, $b \rightarrow j$.

The remaining abbreviations are
\begin{eqnarray*}
A_{L\alpha j}^{(d)}&=& -\frac{m_d}{\sqrt{2}m_W
  cos\beta}U_{j2}^*\left[R_{1\alpha}^{(u)}V_{{\scriptscriptstyle
      CKM}}^{cd}+ R_{3\alpha}^{(u)}V_{{\scriptscriptstyle
      CKM}}^{td}\right] \, \,\,\, (d= b, s) \nn \\
A_{R\alpha j}^{(d)}&=& V_{j1}\left[R_{1\alpha}^{(u)}V_{{\scriptscriptstyle CKM}}^{cd}+R_{3\alpha}^{(u)}V_{{\scriptscriptstyle CKM}}^{td}\right] -\frac{m_c}{\sqrt{2}m_W sin\beta}V_{j2}R_{2\alpha}^{(u)}V_{{\scriptscriptstyle CKM}}^{cd} \nn \\
&-&\frac{m_t}{\sqrt{2}m_W
  sin\beta}V_{j2}R_{4\alpha}^{(u)}V_{{\scriptscriptstyle CKM}}^{td} \,
  \,\,\, (d= b, s)\nn \\
B_{L\alpha a}^{(b)}&=& \sqrt{2}\left[\frac{m_b}{2m_W
  cos\beta}N_{a3}^*R_{3\alpha}^{(d)} +\left[\frac{1}{3}sin\theta_W
  N_{a1}^{'*}-\frac{sin^2\theta_W}{3cos\theta_W}N_{a2}^{'*}\right]R_{4\alpha}^{(d)}\right]\nn
  \\
B_{R\alpha a}^{(b)}&=& 
 \sqrt{2}\left[\left(-\frac{sin\theta_W}{3}N_{a1}^{'}-\frac{1}{cos\theta_W}(\frac{1}{2}-\frac{sin^2\theta_W}{3})N_{a2}^{'}\right)R_{3\alpha}^{(d)}+
\frac{m_b}{2m_W cos\beta}N_{a3}R_{4\alpha}^{(d)}\right]\nn \\
E_{L\alpha a}^{(s)}&=& 
 - \sqrt{2}\left[\left(\frac{sin\theta_W}{3}N_{a1}^{'*}+\frac{1}{cos\theta_W}(\frac{1}{2}-\frac{sin^2\theta_W}{3})N_{a2}^{'*}\right)R_{1\alpha}^{(d)*}-
\frac{m_s}{2m_W cos\beta}N_{a3}^* R_{2\alpha}^{(d)\,*}\right]\nn \\
\end{eqnarray*}
\begin{eqnarray}
E_{R\alpha a}^{(s)}&=& - \sqrt{2}\left[-\frac{m_s}{2m_W
  cos\beta}N_{a3}R_{1\alpha}^{(d)*} +\left(-\frac{1}{3}sin\theta_W
  N_{a1}^{'}+\frac{sin^2\theta_W}{3cos\theta_W}N_{a2}^{'}\right)R_{2\alpha}^{(d)*}\right]\nn
  \\
W_{Lij}^{(x)}&=&\frac{1}{\sqrt{2}}\left(-\sigma_1^{(x)}U_{j2}^*V_{i1}^*+\sigma_2^{(x)}U_{j1}^*V_{i2}^*\right)\nn
  \\
W_{Rij}^{(x)}&=&\frac{1}{\sqrt{2}}\left(-\sigma_1^{(x)}U_{i2}V_{j1}+\sigma_2^{(x)}U_{i1}V_{j2}\right)\nn
  \\
D_{Lab}^{(x)}&=&\frac{1}{2cos\theta_W}\left[(sin\theta_W N_{b1}^*-cos\theta_W N_{b2}^*)(\sigma_1^{(x)}N_{a3}^*+\sigma_2^{(x)}N_{a4}^*)
\right.\nn \\
&+&(sin\theta_W N_{a1}^*-cos\theta_W
  N_{a2}^*)(\sigma_1^{(x)}N_{b3}^*+\sigma_2^{(x)}N_{b4}^*)\left.\right]\nn
  \\
D_{Rab}^{(x)}&=&D_{Lab}^{(x)*}\nn \\
S_{L,q}^{(x)} &=& - \frac{m_q}{2 m_W \cos\beta}\:\sigma_1^{(x)*}\nn \\
S_{R,q}^{(x)} &=&S_{L, q}^{(x)*}
\end{eqnarray}
\[
\textrm{where }
\sigma_1^{(x)} = \begin{pmatrix}
  \sin\alpha& \\
  -\cos\alpha& \\
  i \sin\beta& 
\end{pmatrix},\;\;
\sigma_2^{(x)}=\begin{pmatrix}
 \cos\alpha\\
 \sin\alpha\\
 -i\cos\beta
\end{pmatrix}
\textrm{ for } x=( h_0, H_0, A_0) 
\]
correspondingly, and $N_{a1}^{'}=N_{a1} \cos \theta_W +N_{a2} \sin
\theta_W$, $N_{a2}^{'}=- N_{a1} \sin \theta_W +N_{a2} \cos \theta_W$ .
The  Higgs-squark-squark couplings 
$g_{H_x \tilde q_{\alpha} \tilde q_{\beta}}$ in the physical basis are
given in Appendix A of~\cite{Maria} for the {\it up}-type squarks 
and {\it down}-type squarks independently. The quantities 
$R^{(u)}$, $R^{(d)}$, N, U and V  are the matrices diagonalizing the 
mass matrices of the up squarks, down squarks, neutralinos and
charginos, respectively. 

\section*{Appendix B}

\setcounter{equation}{0}
\renewcommand{\theequation}{B.\arabic{equation}}

In this appendix we give the expressions required to compute the leading contribution to the 
FCHD partial widths in the large SUSY mass limit defined in Sect~\ref{chap.decoupling}, where $M_{S} \sim M_{0} \sim M_{\tilde g} \sim M_1 \sim M_2 \sim \mu 
\sim A  \gg M_{EW}$.
For that purpose we first write the values of the squark, chargino and neutralino masses and their corresponding rotation matrices and then
the formulae for the two- and three-point integrals needed.

\begin{itemize}
\item The expressions for the squark masses and rotation matrices, in the limit of large SUSY mass
parameters and keeping just the leading contribution, are
\begin{eqnarray}
&M_{\tilde q_1}^2 \simeq M_{0}^2 \, \, , \, \,
M_{\tilde q_2}^2 \simeq M_{0}^2 \, , \,
M_{\tilde q_3}^2 \simeq M_{0}^2 (1-\lambda)  \, \, , \, \,
M_{\tilde q_4}^2 \simeq M_{0}^2 (1+\lambda) &
\end{eqnarray}
\begin{eqnarray}
&R_{11}^{(d)} \simeq - R_{12}^{(d)} \simeq R_{43}^{(d)} \simeq R_{44}^{(d)}\simeq 
\frac{m_b}{\sqrt{2}\lambda M_{0}^2} (A- \mu \tan\beta) &\nn \\
&R_{13}^{(d)} \simeq R_{14}^{(d)} \simeq - R_{21}^{(d)} \simeq - R_{22}^{(d)}\simeq
- R_{33}^{(d)} \simeq R_{34}^{(d)} \simeq - R_{41}^{(d)} \simeq R_{42}^{(d)}
\simeq \frac{1}{\sqrt{2}} & \nn \\
&R_{24}^{(d)} \simeq - R_{23}^{(d)} \simeq  R_{31}^{(d)} \simeq R_{32}^{(d)}
\simeq  \frac{m_s}{\sqrt{2}\lambda M_{0}^2} (A-\mu \tan\beta) &   
\end{eqnarray}
and similar results for $R^{(u)}$ just replacing $b \to t$, $s \to c$ and 
$\tan\beta \to \cot\beta$.

\item The expressions for the chargino and neutralino masses and rotation matrices, in the limit of large SUSY mass
parameters can be found in~\cite{rot_neutr}. To leading order these masses are $M_{\tilde \chi_1^-} \simeq M_2$, $M_{\tilde \chi_2^-} \simeq |\mu|$ and $M_{\tilde \chi_1^o} \simeq M_1$, $M_{\tilde \chi_2^o} \simeq M_2$, $ M_{\tilde \chi_3^o}= M_{\tilde \chi_4^o} \simeq |\mu|$,

\item In the large SUSY mass limit the two- and three-point 
one-loop integrals approach their corresponding values at zero external momenta,  $C_{0,11,12}(m_q^2,m_H^2,m_{q'}^2;m_{1}^2, m_{2}^2, m_{3}^2)\simeq C_{0,11,12}(0,0,0;m_{1}^2, m_{2}^2, m_{3}^2)$ and $B_{0,1}(p^2;m_{1}^2,m_{2}^2) \simeq B_0(0;m_{1}^2,m_{2}^2)$, and we write these later as,
\bea
\label{eq.intleading}
&& C_0(0,0,0;m_{1}^2, m_{2}^2, 
m_{3}^2) = -\frac{1}{2 m_2^2} f_1 (R_{1}, R_{2})  \nn\\
&& C_{11}(0,0,0;m_{1}^2, m_{2}^2, 
m_{3}^2) = \frac{1}{3 m_2^2} f_{10} (R_{1},R_{2}) \nn \\
&& C_{12}(0,0,0;m_{1}^2, m_{2}^2, 
m_{3}^2) = \frac{1}{6 m_2^2} f_{13} (R_{1},R_{2})  \nn\\
&&B_0(0;m_{1}^2,m_{2}^2)
    =  \Delta - log\frac{m_{2}^2}{\mu_0^2} + g_1 (R_{1}) \nn \\ 
&&B_1(0;m_{1}^2,m_{2}^2)
    = - \frac{1}{2} \Delta + \frac{1}{2} log\frac{m_{2}^2}{\mu_0^2}-g_1 (R_{1})  - g_2 (R_{1}) , 
\label{eq.expintegrals}
\eea
where $m_1$, $m_2$, and $m_3$, represent generically the masses of the different particles inside the loops, $R_{1} = m_{1}/m_{2}$, $R_{2} = m_{1}/m_{3}$  and the explicit
 expressions for the functions 
$f_i$ and $g_i$ can be found in~\cite{ourHtb}.
For the simplest case, where $R_{1}=R_{2}=1$, they are 
$f_1(1,1)=f_{10}(1,1)=f_{13}(1,1)=1$ and $g_1(1)=g_2(1)=0$.
\end{itemize}

\section*{Appendix C}

\setcounter{equation}{0}
\renewcommand{\theequation}{C.\arabic{equation}}
In this appendix we present the expansions of the SUSY-EW contributions to 
the form factors in inverse
powers of $M_S$ where $M_{S} \sim M_{0} \sim M_{\tilde g} \sim M_1 \sim M_2 \sim \mu \sim A  \gg M_{EW}$ and keeping just the leading contribution. The following results are valid for the most general case of arbitrary $\lambda$, $\gamma$, $\eta$ and $\tan \beta$, where $\gamma$ and $\eta$ are defined by, $M_1=\gamma M_{\tilde g}$ and $M_2=\eta M_{\tilde g}$. The form factor is,
\begin{eqnarray}\hspace*{-0.9cm}
&&\hspace*{-1.5cm}F_{L}^{(x)} = F_{L_{a_C}}^{(x)}+ F_{L_{b_C}}^{(x)}
+F_{L_{c_C+d_C}}^{(x)}+F_{L_{a_N}}^{(x)}+F_{L_{b_N}}^{(x)}
+F_{L_{c_N+d_N}}^{(x)}\,,
\label{formfactorLdecoupequalescalebs1}
\end{eqnarray}
where the contributions from the different diagrams 
($a_{C,N}$, $b_{C,N}$, $c_{C,N}$, and $d_{C,N}$) are given respectively by,
\begin{eqnarray}\hspace*{-0.9cm}
&&\hspace*{-1.5cm}F_{L_{a_C}}^{(x)} = \frac{g^2}{16 \pi^2} \frac{m_b}{2 m_W \cos \beta}\left[\frac{\sigma_1^{(x)*}}{4}\left[\left(V_{{\scriptscriptstyle CKM}}^{cb} V_{{\scriptscriptstyle CKM}}^{cs} + V_{{\scriptscriptstyle CKM}}^{tb} V_{{\scriptscriptstyle CKM}}^{ts}\right)\left(G(\lambda,\eta)+ \lambda S(\lambda,\eta)\right)
\right. \right. \nn \\
&& \left. 
+\left( V_{{\scriptscriptstyle CKM}}^{cb} V_{{\scriptscriptstyle CKM}}^{ts} + V_{{\scriptscriptstyle CKM}}^{tb} V_{{\scriptscriptstyle CKM}}^{cs}\right) \left(S(\lambda,\eta) + \lambda G(\lambda,\eta)\right)\right] \nn \\
&& 
- \frac{M_2 \mu}{M_S^2} \frac{\sigma_2^{(x)}}{4} \left[\left( V_{{\scriptscriptstyle CKM}}^{cb} V_{{\scriptscriptstyle CKM}}^{cs} + V_{{\scriptscriptstyle CKM}}^{tb} V_{{\scriptscriptstyle CKM}}^{ts}\right) G(\lambda,\eta)
\right. \nn \\
&& \left. \left.
+\left( V_{{\scriptscriptstyle CKM}}^{cb} V_{{\scriptscriptstyle CKM}}^{ts} + V_{{\scriptscriptstyle CKM}}^{tb} V_{{\scriptscriptstyle CKM}}^{cs}\right) S(\lambda,\eta)\right] 
{\phantom{\frac{\sigma_1^{(x)*}}{4}}}\hspace*{-1.0cm}\right]\\
\label{formfactorLdecoupequalescalebs2}
\nn \\
&&\hspace*{-1.5cm}F_{L_{b_C}}^{(x)} = \frac{g^2}{16 \pi^2} \frac{m_b}{2 m_W \cos \beta}
\frac{\mu (A \sigma_2^{(x)}+\mu \sigma_1^{(x)*})}{M_S^2}\,\frac{1}{8 m_W^2 \ \sin^2 \beta}
  \left[\left( V_{{\scriptscriptstyle CKM}}^{cb} V_{{\scriptscriptstyle CKM}}^{ts} m_t^2 
\right. \right.\nn \\
&&\left.\left. 
+ V_{{\scriptscriptstyle CKM}}^{tb} V_{{\scriptscriptstyle CKM}}^{cs} m_c^2 \right) F(\lambda) 
+ \left( V_{{\scriptscriptstyle CKM}}^{cb} V_{{\scriptscriptstyle CKM}}^{cs} m_c^2 + V_{{\scriptscriptstyle CKM}}^{tb}
  V_{{\scriptscriptstyle CKM}}^{ts} m_t^2 \right) J(\lambda)\right]\\
\label{formfactorLdecoupequalescalebs3}
\nn \\
&&\hspace*{-1.9cm}F_{L_{c_C+d_C}}^{(x)} = -\frac{g^2}{16 \pi^2} \frac{m_b}{2 m_W \cos \beta}\sigma_1^{(x)*}\left[\frac{\mu \left(\mu - A \tan \beta\right)}{4 \sin 2\beta m_W^2 \tan \beta  M_S^2}
\left[\left( V_{{\scriptscriptstyle CKM}}^{tb} V_{{\scriptscriptstyle CKM}}^{cs} m_c^2 
\right. \right. \right. \nn \\
&& \left. \left.
+ V_{{\scriptscriptstyle CKM}}^{cb} V_{{\scriptscriptstyle CKM}}^{ts} m_t^2\right) F(\lambda)
+\left( V_{{\scriptscriptstyle CKM}}^{cb} V_{{\scriptscriptstyle CKM}}^{cs} m_c^2  + V_{{\scriptscriptstyle CKM}}^{tb} V_{{\scriptscriptstyle CKM}}^{ts} m_t^2\right) J(\lambda)\right]
 \nn \\
&& 
- \frac{1}{4 \cos \beta (M_2^2-\mu^2)} \left[\mu \left( M_2 \sin \beta + \mu \cos \beta \right)\left[\left( V_{{\scriptscriptstyle CKM}}^{cb} V_{{\scriptscriptstyle CKM}}^{cs} + V_{{\scriptscriptstyle CKM}}^{tb} V_{{\scriptscriptstyle CKM}}^{ts}\right)F(\lambda)\lambda
\right. \right.\nn \\
&& \left. 
+\left( V_{{\scriptscriptstyle CKM}}^{cb} V_{{\scriptscriptstyle CKM}}^{ts} + V_{{\scriptscriptstyle CKM}}^{tb} V_{{\scriptscriptstyle CKM}}^{cs}\right) J(\lambda)\lambda\right] 
 \nn \\
&& 
+ 2 M_2 \left( M_2 \cos \beta + \mu \sin \beta \right)\left[\left( V_{{\scriptscriptstyle CKM}}^{cb} V_{{\scriptscriptstyle CKM}}^{cs} + V_{{\scriptscriptstyle CKM}}^{tb} V_{{\scriptscriptstyle CKM}}^{ts}\right)L(\lambda,\eta)
\right. \nn \\
&& \left. \left. \left.
+\left( V_{{\scriptscriptstyle CKM}}^{cb} V_{{\scriptscriptstyle
      CKM}}^{ts} + V_{{\scriptscriptstyle CKM}}^{tb}
  V_{{\scriptscriptstyle CKM}}^{cs}\right)P(\lambda,\eta)\right]\right]
{\phantom{\frac{1}{4 \mu}}}\hspace*{-0.6cm}\right]
\label{formfactorLdecoupequalescalebs4}
\end{eqnarray}\vspace*{-0.2cm}
\begin{eqnarray}\hspace*{-0.9cm}
&&\hspace*{-1.5cm}F_{L_{a_N}}^{(x)} = \frac{g^2}{16 \pi^2}\frac{m_b }{2 m_W \cos \beta}\left[ \frac{\tan ^2 \theta_W}{24} \left( \left(S(\lambda,\gamma)+\lambda G(\lambda,\gamma)\right) \sigma_1^{(x)*}- \frac{M_1 \mu}{M_S^2} \sigma_2^{(x)} S(\lambda,\gamma)\right)\right.\nn \\
&& \left. + 
 \frac{1}{8} \left( \left(S(\lambda,\gamma)+\lambda
     G(\lambda,\gamma)\right) \sigma_1^{(x)*}- \frac{M_2 \mu}{M_S^2}
   \sigma_2^{(x)} S(\lambda,\eta)\right)\right]\\
\nn \\
&&\hspace*{-1.5cm}F_{L_{b_N}}^{(x)} = - \frac{g^2}{16 \pi^2}\frac{m_b}{2 m_W \cos \beta}(A \sigma_1^{(x)*} + \mu \sigma_2^{(x)}) \left[ \frac{M_1}{2 M_S^2} \frac{\tan ^2 \theta_W}{18} S(\lambda,\gamma) \right]
\label{formfactorLdecoupequalescalebs6}
\end{eqnarray}

\begin{eqnarray}\hspace*{-0.9cm}
&&\hspace*{-1.5cm}F_{L_{c_N+d_N}}^{(x)} = - \frac{g^2}{16 \pi^2} \frac{m_b}{2 m_W \cos \beta} \sigma_1^{(x)*}\left[\frac{\tan ^2 \theta_W}{12} M_1\left( - \frac{1}{\cos \beta (M_1^2-\mu^2)}(M_1 \cos \beta + \mu \sin \beta)P(\lambda,\gamma) 
\right. \right. \nn \\
&& \left. + \frac{2}{3} \frac{\left( A - \mu \tan \beta \right)}{M_S^2}K(\lambda,\gamma)  \right)-
M_2 \left( \frac{1}{4 \cos \beta (M_2^2-\mu^2)}(M_2 \cos \beta + \mu \sin \beta)P(\lambda,\eta) \right)  \nn \\
&& \left. - \mu \left(\frac{1}{24 \cos \beta} \left[ \frac{3}{(M_2^2-\mu^2)}(M_2 \sin \beta + \mu \cos \beta) +\frac{\tan ^2 \theta_W}{(M_1^2-\mu^2)}(M_1 \sin \beta + \mu \cos \beta) \right]J(\lambda)\lambda  \right)
\right] \nn \\
\label{formfactorLdecoupequalescalebs7}
\end{eqnarray}

$F_{R}^{(x)}$ can be obtained by replacing all ($L \leftrightarrow R$), 
$m_b\leftrightarrow m_s $, $m_t\leftrightarrow m_c $, and taking the
complex conjugate. The previous result of 
eq.~(\ref{formfactorLdecoupequalescalebs1}) is valid for all $m_A$ and 
$\tan \beta$ values and keep all the involved quark masses, 
$m_t$, $m_b$, $m_c$ and $m_s$, different from zero.
The different functions of $\lambda$ that appear in the above expressions and their behaviour in the $\eta \to 1$ and $\gamma \to 1$ limits are given by, 
\begin{eqnarray}
F(\lambda) &=& \frac{2}{\lambda^2} [(\lambda +1)\ln(\lambda+1) + 
(\lambda -1)\ln(1-\lambda) - 2\lambda]; \, \,\lim_{\lambda \to 0} F(\lambda) \simeq -\frac{2 \lambda}{3}-\frac{\lambda^3}{5},
\nn\\
J(\lambda) &=& \frac{2}{\lambda^2} [(\lambda +1)\ln(\lambda+1) -
(\lambda -1)\ln(1-\lambda)]; \, \, \lim_{\lambda \to 0} J(\lambda) \simeq  2 + \frac{\lambda^2}{3},
\nn\\
S(\lambda,\eta) &=& f_1 \left(\sqrt{1+\lambda},\frac{\sqrt{1+\lambda}}{\eta}\right)-f_1 \left(\sqrt{1-\lambda},\frac{\sqrt{1-\lambda}}{\eta}\right); \, \, S(\lambda,1)=F(\lambda),
\nn\\
G(\lambda,\eta) &=& f_1 \left(\sqrt{1+\lambda},\frac{\sqrt{1+\lambda}}{\eta}\right)+f_1 \left(\sqrt{1-\lambda},\frac{\sqrt{1-\lambda}}{\eta}\right); \, \,
G(\lambda,1)=J(\lambda),\nn\\
L(\lambda,\eta)& =&-\ln(1+\lambda)-\ln(1-\lambda)+g_1\left(\frac{\eta}{\sqrt{1+\lambda}}\right)+g_1\left(\frac{\eta}{\sqrt{1-\lambda}}\right); \, \, L(\lambda,1)=-\frac{F(\lambda) \lambda}{2},
\nn\\ 
P(\lambda,\eta) &=&-\ln(1+\lambda)+\ln(1-\lambda)+g_1\left(\frac{\eta}{\sqrt{1+\lambda}}\right)-g_1\left(\frac{\eta}{\sqrt{1-\lambda}}\right); \, \, P(\lambda,1)=-\frac{J(\lambda) \lambda}{2},
\nn\\
K(\lambda,\eta) &=& \frac{1}{\lambda}\left(L(\lambda,\eta)-2g_1(\eta) \right); \, \, K(\lambda,1)=-\frac{F(\lambda)}{2}\,.
\label{lambdafunctions}
\end{eqnarray}\vspace*{0.1cm}

 In the following we present our results for equal SUSY mass parameters. The
 contributions from charginos are,
\begin{eqnarray}\hspace*{-0.9cm}
&&\hspace*{-2.5cm}F_{L_{a_C}}^{(x)}= \frac{g^2}{16 \pi^2} \frac{m_b}{2 m_W \cos \beta}\left[\frac{\sigma_1^{(x)*}}{4}\left[\left( V_{{\scriptscriptstyle CKM}}^{cb} V_{{\scriptscriptstyle CKM}}^{cs} + V_{{\scriptscriptstyle CKM}}^{tb} V_{{\scriptscriptstyle CKM}}^{ts}\right) \left(J(\lambda)+\lambda F(\lambda)\right)
\right. \right. \nn \\
&& \left. 
+\left( V_{{\scriptscriptstyle CKM}}^{cb} V_{{\scriptscriptstyle CKM}}^{ts} + V_{{\scriptscriptstyle CKM}}^{tb} V_{{\scriptscriptstyle CKM}}^{cs}\right) \left(F(\lambda)+ \lambda J(\lambda)\right)\right] 
 \nn \\
&& 
-  \frac{\sigma_2^{(x)}}{4} \left[\left( V_{{\scriptscriptstyle CKM}}^{cb} V_{{\scriptscriptstyle CKM}}^{cs} + V_{{\scriptscriptstyle CKM}}^{tb} V_{{\scriptscriptstyle CKM}}^{ts}\right) J(\lambda)
\right. \nn \\
&& \left. \left.
+\left( V_{{\scriptscriptstyle CKM}}^{cb} V_{{\scriptscriptstyle CKM}}^{ts} + V_{{\scriptscriptstyle CKM}}^{tb} V_{{\scriptscriptstyle CKM}}^{cs}\right) F(\lambda)\right] 
{\phantom{\frac{\sigma_1^{(x)*}}{4}}}\hspace*{-1.0cm}\right]
\label{formfactorLdecoupequalescalebs9}
\end{eqnarray}
\begin{eqnarray}\hspace*{-0.9cm}
&&\hspace*{-2.5cm}F_{L_{b_C}}^{(x)} = \frac{g^2}{16 \pi^2} \frac{m_b}{2
  m_W \cos \beta} (\sigma_2^{(x)}+ \sigma_1^{(x)*})\,\frac{1}{8 m_W^2
  \sin^2 \beta} \left[\left( V_{{\scriptscriptstyle CKM}}^{cb} 
V_{{\scriptscriptstyle CKM}}^{ts} m_t^2 \right. \right. \nn \\
&& \left. \left.
 + V_{{\scriptscriptstyle CKM}}^{tb} V_{{\scriptscriptstyle CKM}}^{cs} m_c^2\right) F(\lambda) + \left( V_{{\scriptscriptstyle CKM}}^{cb} V_{{\scriptscriptstyle CKM}}^{cs} m_c^2 + V_{{\scriptscriptstyle CKM}}^{tb} V_{{\scriptscriptstyle CKM}}^{ts} m_t^2\right) J(\lambda)\right]
\label{formfactorLdecoupequalescalebs10}
\end{eqnarray}

\begin{eqnarray}\hspace*{-0.9cm}
&&\hspace*{-2.5cm}F_{L_{c_C+d_C}}^{(x)} = -\frac{g^2}{16 \pi^2} \frac{m_b}{2 m_W \cos \beta}\sigma_1^{(x)*}\left[\frac{\left(1 -  \tan \beta\right)}{8 m_W^2 \sin^2\beta}
\left[\left( V_{{\scriptscriptstyle CKM}}^{tb} V_{{\scriptscriptstyle CKM}}^{cs} m_c^2 
\right. \right. \right. \nn \\
&& \left. \left.
+ V_{{\scriptscriptstyle CKM}}^{cb} V_{{\scriptscriptstyle CKM}}^{ts} m_t^2\right) F(\lambda)+\left( V_{{\scriptscriptstyle CKM}}^{cb} V_{{\scriptscriptstyle CKM}}^{cs} m_c^2  + V_{{\scriptscriptstyle CKM}}^{tb} V_{{\scriptscriptstyle CKM}}^{ts} m_t^2\right) J(\lambda)\right]
 \nn \\
&& 
+ \frac{1}{4} \left[\left(V_{{\scriptscriptstyle CKM}}^{cb} V_{{\scriptscriptstyle CKM}}^{cs} + V_{{\scriptscriptstyle CKM}}^{tb} V_{{\scriptscriptstyle CKM}}^{ts}\right)\left(\lambda F(\lambda)+J(\lambda)\left(1+\tan\beta\right)\right)
\right.\nn \\
&& \left. \left.
+\left( V_{{\scriptscriptstyle CKM}}^{cb} V_{{\scriptscriptstyle CKM}}^{ts} + V_{{\scriptscriptstyle CKM}}^{tb} V_{{\scriptscriptstyle CKM}}^{cs}\right) \left(\lambda J(\lambda)+F(\lambda)\left(1+\tan\beta\right)\right)\right] 
{\phantom{\frac{\left(1 -  \tan \beta\right)}{8 m_W^2}}}\hspace*{-2.0cm}\right]
\label{formfactorLdecoupequalescalebs11}
\end{eqnarray}

For neutralinos we get,
\begin{eqnarray}\hspace*{-0.9cm}
&&\hspace*{-2.5cm}F_{L_{a_N}}^{(x)}= \frac{g^2}{16 \pi^2} \frac{m_b}{2 m_W \cos \beta}\left[\frac{1}{8}\left(1+\frac{\tan ^2\theta_W}{3}\right) \left(\sigma_1^{(x)*}\left(F(\lambda)+\lambda J(\lambda)\right)- \sigma_2^{(x)}F(\lambda)\right)\right]\\
\label{formfactorLdecoupequalescalebs12}
\nn \\
&&\hspace*{-2.5cm}F_{L_{b_N}}^{(x)} =-\frac{g^2}{16 \pi^2}\frac{m_b}{2 m_W \cos \beta}\left[\frac{\tan ^2\theta_W}{36}\left( \sigma_1^{(x)*}+ \sigma_2^{(x)}\right)
\right]F(\lambda)\\
\label{formfactorLdecoupequalescalebs13}
\nn \\
&&\hspace*{-2.5cm}F_{L_{c_N+d_N}}^{(x)} = -\frac{g^2}{16 \pi^2} \frac{m_b}{2 m_W \cos \beta}\sigma_1^{(x)*}\left[\frac{1}{8}\left(\frac{\tan ^2\theta_W}{3}+1\right) \left(\lambda J(\lambda)+F(\lambda)\left(1+\tan\beta\right)\right) 
\right. \nn \\
&& \left. -\frac{\tan ^2\theta_W}{36} \left(1 - \tan \beta\right) F(\lambda)
\right]
\label{formfactorLdecoupequalescalebs14}
\end{eqnarray} 
 

\end{document}